# Crossover from quantum correlation to hot-carrier transport in scattering-tolerant 2D transistors

Debottam Daw[1,2], Houcine Bouzid[2], Sung-Gyu Lee[2], Wujoon Cha[2], Ki Kang Kim[2,3], Min-kyu Joo[4,5], Yan Wang[6], Manish Chhowalla[6], and Young Hee Lee[1,2,7*]

[1]Center for Low-Dimensional Quantum Materials, Hubei University of Technology, Wuhan 430062, P. R. China.

[2]Center for Integrated Nanostructure Physics, Sungkyunkwan University, Suwon 16419, Republic of Korea.

[3]Department of Energy Science, Sungkyunkwan University, Suwon 16419, Republic of Korea.

[4]Department of Applied Physics, Sookmyung Women's University, Seoul, Republic of Korea.

[5]Institute of Advanced Materials and Systems, Sookmyung Women's University, Seoul, Republic of Korea.

[6]Department of Materials Science & Metallurgy, University of Cambridge, 27 Charles Babbage Road, Cambridge CB3 0FS, UK.

[7]School of Materials Science and Engineering, Peking University, Beijing 100871, P. R. China.

*Email: leeyoung@skku.edu

## Abstract

**Quantum correlation and hot-carrier transport represent two fundamentally distinct regimes of electronic conduction, rarely accessible within the same device. Here, we report a state-of-the-art monolayer transition metal dichalcogenides transistor architecture on a ferroelectric substrate that enables this crossover by leveraging the strong dielectric screening and in-plane gate control. At cryogenic temperatures, the devices exhibit reproducible quasi-periodic current fluctuations, consistent with an emergent potential landscape driven by electron-electron interactions at low carrier densities. As the temperature increases, this correlated potential profile thermally dissolves and transport is dominated by the lateral gate-field that drives the carriers with high kinetic energy. These hot-carriers can efficiently surmount the scattering events, exhibiting a record-high room-temperature electron mobility of ~4,800 $cm^2V^{-1}s^{-1}$ and a maximum on-current ~0.5 mA/µm, surpassing traditional FETs in key performance metrics. These findings establish a unified**

**approach for probing intermediate mesoscopic orders, while advancing the transistor performance limits in scalable 2D transistors.**

## Main

Carrier transport in two-dimensional (2D) layered semiconductors is typically limited by high intrinsic Coulomb scattering and strong electron-phonon interactions. These scattering mechanisms restrict both the carrier mobility and mask the effects of electron-electron interaction-driven delicate phases. While the advanced transistor designs have achieved incremental performance gains, a unified strategy that captures the underlying transport mechanisms across a wide temperature range while pushing the performance limits has remained largely unexplored. In atomically thin disordered 2D systems, low-temperature conduction is often dominated by charge trapping, thermally activated hopping, or classical percolation(*1*, *2*). However, under certain conditions if the carrier density is sufficiently low, electron-electron interactions can reshape the potential landscape, leading to emergent correlation that deviates from conventional transport(*3*). At elevated temperatures, the increased phonon and carrier-carrier scatterings further complicate the transport by suppressing carrier mobility(*4*, *5*). Numerous strategies to enhance carrier mobility have been pursued including contact engineering(*6*, *7*), strain-induced band-structure tuning(*8*–*10*), reducing electron-phonon coupling through surface morphology control(*11*, *12*), optimizing dielectric screening(*13*, *14*), and minimizing surface scattering with dielectric encapsulation(*15*, *16*). Nevertheless, marginal improvements have been achieved to date. Moreover, these opposing transport regimes have traditionally required separate experimental platforms, hindering their broader applicability in scalable nanoelectronics.

Here, we report a transistor architecture that bridges these disparate transport regimes by enabling a crossover from correlated current fluctuations at cryogenic temperatures to hot-carrier dominated high-mobility transport at room temperature. The device consists of semiconducting $MX_2$ monolayer integrated on an insulating ferroelectric substrate, featuring a unique remote side-gate configuration. The ferroelectric substrate with in-plane polarization acts as the side-gate dielectric and significantly enhances the polarization of free carriers within the channel. The side-gate electrode, positioned remotely (typically 2-5 μm) from the drain terminal, establishes a robust in-plane electric field along the channel. This configuration allows the carriers in the channel to gain kinetic energy from the induced gate field. These high-energy carriers, referred to as hot carriers, are capable of surpassing the scattering events that typically impede mobility. Our results demonstrate remarkable performance in both n-type and p-type HC-FETs, achieving high current densities exceeding ~0.5 mA/μm, alongside peak room-temperature mobilities over ~4,800 $cm^2V^{-1}s^{-1}$. Additionally, these devices exhibit an impressive on-off current ratio of ~$10^8$, and near-ideal, hysteresis-free subthreshold swing (SS) despite the presence of ferroelectric substrate. At cryogenic temperatures, we observe well-defined, reproducible conductance fluctuations that show systematic thermal smearing with increasing temperature. Resonant transport modeling and self-consistent electrostatic simulations indicate that long-range electron-electron interactions shape an effective quasi-periodic potential landscape even in the absence of classical confinement.

**Strategies for mitigating scattering events**

Figure 1 introduces the operational principle of the HC-FET, which integrates monolayer $MoS_2$ on a lead magnesium niobate-led titanate (PMN-PT) (001) substrate adopting side-gate configuration (Fig. 1A, schematics and Supplementary materials section 1). This device combines two physical effects that together govern its transport behavior: (i) in-plane field-induced carrier acceleration and (ii) screening of charge impurities by the ferroelectric substrate.

PMN-PT forms stable in-plane ferroelectric micro-domains(*17*) below its Curie temperature ($T_C$ = 360 K)(*18*, *19*) enabling substantial electrostatic coupling between gate and the channel. In side-gate configuration, above the threshold voltage, a relatively small number of carriers acquire excess kinetic energy from the cooperative effect of drain field and in-plane gate polarization-induced field (Fig. 1A, right). These carriers remain within the same conduction band but are accelerated into higher-momentum states, allowing them to efficiently surmount scattering centers in the band edge. This process reflects a non-equilibrium carrier distribution above the band edge rather than intra-band transition and is analogous to hot-electron dynamics observed in high-field transport(*20*–*22*).

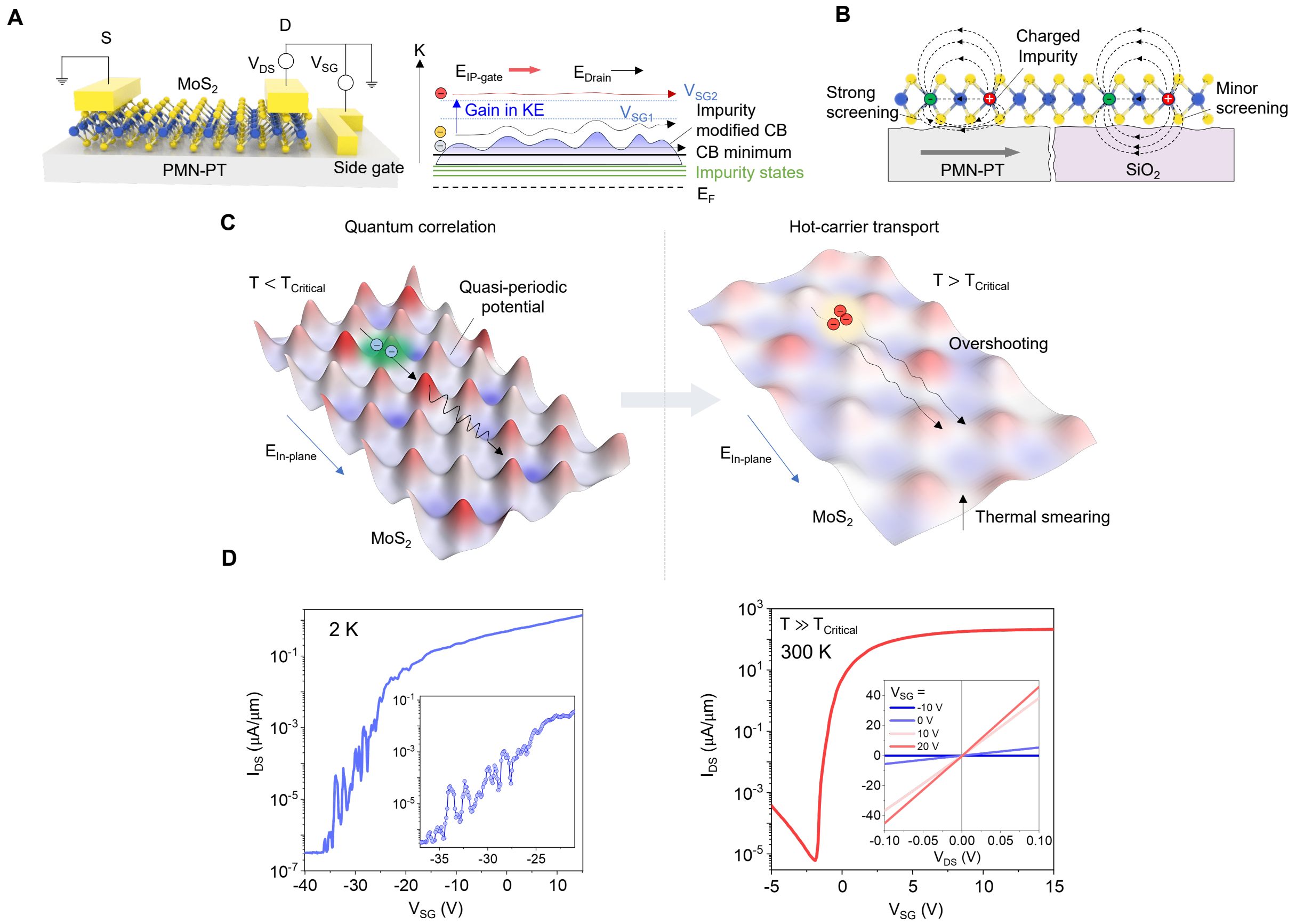

**Fig. 1. Schematics and working principle of HC-FET**. (**A**) Schematic device of $MoS_2$ HC-FET on PMN-PT substrate. The presence of strong in-plane polarization with side-gate leads to unconventional band modulation and hence hot-carrier generation above scattering centers (right). (**B**) Comparison of dielectric screening between PMN-PT and conventional $SiO_2$. The high dielectric constant of PMN-PT enables stronger screening of charged impurities. (**C**) Conceptual outlook of emergent electron-electron interaction-stabilized correlation at low temperature and low carrier density (left). At high temperatures, thermal smearing and in-plane field induced acceleration enable hot-carrier transport (right). (**D**) Experimental signature (transfer characteristic) of correlation induced current fluctuation at 2 K (left). The inset shows high-resolution data revealing discrete fluctuation peaks. Hot-carrier enabled smooth transfer characteristics at room temperature. Linear output characteristics (inset) confirming ohmic contact behavior without spurious contact-induced artifacts.

Simultaneously, the PMN-PT substrate plays a decisive role in minimizing the scattering from charged impurities. In atomically thin TMDs, like $MoS_2$, the impurity potential is spatially extended and predominantly influenced by the dielectric environment(*23*). High permittivity of the PMN-PT substrate significantly reduces the scattering potential through efficient charge screening (Fig. 1B), as shown by electrostatic simulations (fig. S1, Supplementary materials section 2.1). Furthermore, in-plane polarization of the PMN-PT substrate enhances the polarizability of free carriers, which further suppresses scatterings from residual disorders (see Supplementary materials section 2.2 and fig. S2). The resulting increase in screening supports efficient high-temperature transport and emergence of delicate low-temperature phenomena(*15*).

The relative impact of these mechanisms varies across two regimes: Quantum correlation and hot-carrier transport (Fig. 1C). At cryogenic temperatures, particularly in the subthreshold regime where carrier density is low, the suppression of impurities and phonons, allows the long-range electron-electron interactions to take over the kinetic energy of the carriers, leading to formation of quasi-periodic potential landscape across the channel(*24*). As the temperature increases, this potential landscape is gradually destabilized by thermal broadening, and field-driven carrier acceleration becomes the dominant transport mechanism.

The crossover is directly manifested in the measured transfer curves (Fig. 1D). At 2 K, the device exhibits reproducible, quasi-periodic current fluctuations in the subthreshold region, while at 300 K, the same device shows clean, hysteresis-free ohmic (inset) switching behavior, characteristic of high-performance FETs. The coexistence of these two regimes within a single device raises two central questions: What governs the microscopic mechanism of hot-carrier transport in this geometry and how does electron-electron interaction, in the absence of moiré engineering or intentional confinement, give rise to correlated fluctuation patterns? These questions form the basis of the analyses presented in the following sections.

**Reduced scattering *via* in-plane field**

Figure 2A presents the transfer characteristics (see Methods and fig. S3 for fabrication details) of n-type $MoS_2$ HC-FET (inset shows the optical micrograph) at room temperature. In typical back-gate FETs (BG-FET), the application of a drain voltage shifts the Fermi level from its equilibrium position(*16*), resulting in lateral band bending which creates an electric field (fig. S4 a) responsible for carrier-drift along the channel. Meanwhile, the gate bias modulates the channel barrier height thereby injecting or blocking majority carriers to achieve transistor switching. In this configuration, however, the gate field plays no active role in driving carriers in the channel. In contrast, a side-gate bias can help driving carriers in conjunction with the drain voltage by establishing an electrostatic field along the channel without physical overlapping. Conventional dielectric substrates like $SiO_2$ are ineffective for such a side-gate configuration, owing to limited electrostatic control over the channel (fig. S4 b). The HC-FET addresses these challenges by adopting side-gate geometry on ferroelectric PMN-PT substrate (fig. S4 c), offering over an order of higher current density of ~0.5 mA/μm compared to the conventional vertical gate-FETs of similar dimensions. Ohmic contact was achieved using semi-metal Bi/Au, as further evidenced by the direct tunnelling behavior (fig. S5)(*7*). We confirmed the band modulation in monolayer-$MoS_2$ HC-FET using Kelvin probe force microscopy (fig. S6, and Supplementary materials section 2.3).

To demonstrate the effectiveness of the side-gate configuration over traditional back (or top) gate, we utilized finite element method (FEM) simulations (fig. S7-S10). The associated potential and field distributions clearly point towards significant in-plane field-modulation (table S1) and electrostatic coupling (table S2) in side-gate configuration compared to that of back-gate. When the gate-induced field aligns with the drain-to-source field, the carriers are expected to gain higher local drift-velocity in the channel. Although the elevated drift-velocity is not directly related to mobility improvement, carriers with higher kinetic energy are less likely to lose energy due to fewer available phonon modes, allowing them to traverse the channel with reduced interactions with scattering centers.

**Transport measurements at room temperature**

The field-effect mobility and the carrier density were extracted from the linear region of the transfer characteristics following, $\mu_{FE} = \frac{\partial I_{DS}}{\partial V_{GS}} \left( \frac{L}{W C_{\text{in}} V_{DS}} \right)$ and $n = C_{\text{in}} \left( V_{GS} - V_{th} \right) / e$(*16*). Both the equations hold true for the linear charge dependence on the gate over-drive bias. The required capacitance values for mobility calculation were experimentally measured (~40 $nF/cm^2$ on average) for HC-FETs (fig. S11). The maximum room-temperature mobility extracted for the featured n-type HC-FET (D-14) was ~4,770 $cm^2V^{-1}s^{-1}$ at an on-current level of 70 μA, which is the highest in any TMD FETs to date, (Fig. 2B). This is well contrasted with a typical low mobility of ~97 $cm^2V^{-1}s^{-1}$ (Fig. 2C) at an on-current level below 10 μA (inset) observed in conventional $MoS_2/SiO_2$ BG-FET(*7*, *25*). We further assessed (using FEM simulation) the impact of change in gate configuration on FET mobility by implementing Lombardi and Caughey-Thomas models accounting for relevant scattering events (fig. S12 and Supplementary materials section 2.4)(*26*, *27*). Nevertheless, given the reference input mobility, side-gate FETs always exhibit greater resilience to scattering, leading to improved mobility compared to the standard

BG-FETs. This supports the experimental observation on mobility enhancement by efficient screening of scattering events. Reproducibility was demonstrated across 21 different n-type HC-FETs with mobilities ranging from 820-4,770 $cm^2V^{-1}s^{-1}$ with an average of 2,200±10 $cm^2V^{-1}s^{-1}$ (fig. S13a,b). Both two and four-probe measurements (*e.g.*, device D-9 in fig. S13 c) confirm consistent mobility values, attributed to the Ohmic nature of the Bi contacts(*28*), ruling out any contact-induced effects on mobility estimation. The performance variations in different devices originate from the unexpected residues below the contacts, formed during the device fabrication process. Furthermore, the mobility (*e.g.* device D-14) remained stable for eight months (fig. S14), indicating reliable performance under ambient conditions.

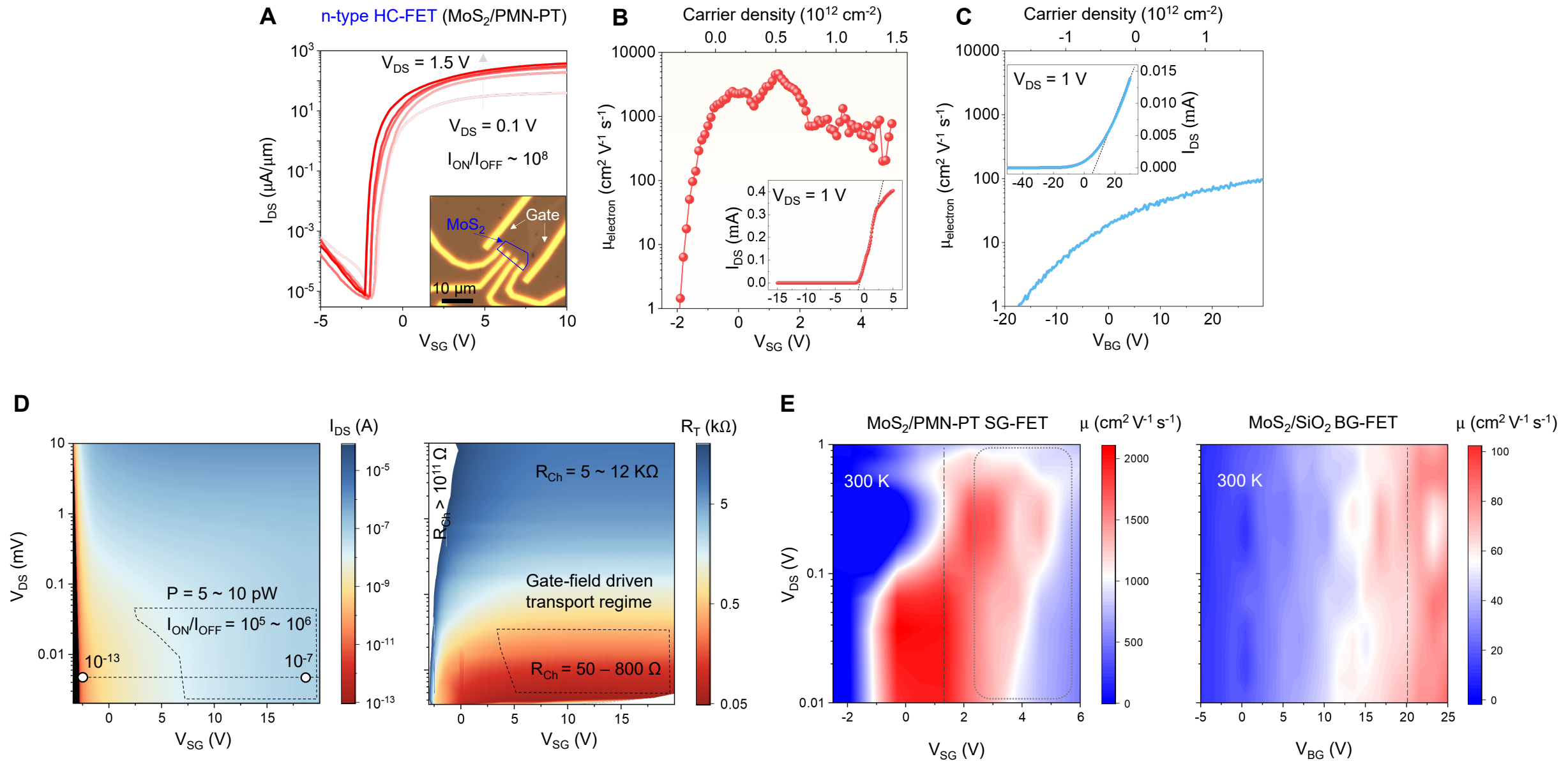

**Fig. 2. Room-temperature performance analysis of HC-FETs**. (**A-B**) Transfer characteristics at different $V_{DS}$ with optical micrographs of the device (channel length $L$ = 1 μm) in the inset (**A**) and two-probe FET mobility for n-type $MoS_2$ HC-FET with linear scale $I_{DS}$-$V_{SG}$ plot (inset) (**B**). (**C**) Gate-dependent mobility of $MoS_2/SiO_2$ BG-FET with the linear scale output characteristics measured under similar experimental configurations. (**D**) Current mapping of HC-FET under low supply voltage reveals efficient transport under influence of side-gate gate (left). The device can operate at picowatt power level while having minimal resistance (right). (**E**) Contour plot of the room temperature mobility under varying drain voltage and gate bias for $MoS_2$ HC-FET (left) and BG-FET (right).

The HC-FET exhibited a subthreshold swing (SS) of 68 mV/dec, approaching the thermal limit (fig. S15), well contrasted with a large value of ~2,410 mV/dec in BG-FET. Such a nearly ideal SS in HC-FET could be attributed to the influence of ferroelectric substrate(*29*). Notably, HC-FET exhibited the negligible hysteresis (fig. S16 a,b), avoiding common pitfalls seen in typical ferroelectric-FETs(*30*). The nominal hysteresis in our HC-FET may result from residual defects or trap charges, as also seen in conventional FETs (fig. S16 c). We successfully extended our concept of hot carriers from n-type HC-FETs to p-type monolayer $WSe_2$-based HC-FETs (fig. S17 a,b). The highest hole mobility of 1,847 $cm^2V^{-1}s^{-1}$ for p-type HC-FET was derived from the linear region of the transfer characteristics (inset) at an on-current level of 0.1 mA. Similar performance was reproducibly demonstrated with another $WSe_2$ HC-FET (fig. S17 c). Both p- and n-type HC-FETs showcased superior performances, underscoring their potential to significantly enhance the capabilities of next-generation CMOS technology.

**Impact of screening on mobility enhancement**

To assess the role of the side-gate on carrier transport under minimum drain influence, we characterized the HC-FET at ultra-low supply voltage window (Fig. 2D), where typical FETs exhibit gate-leakage limited conduction (~pA). Remarkably, even at drain bias as low as few tens of microvolts, the device sustains sizable on-currents (~0.01-0.1 μA) with a minimal resistance of ~500 Ω - 1 kΩ, indicating the ability of the in-plane gate field to influence the carrier motion independently of the drain field. However, at high drain bias, this effect becomes less prominent due to the increasing dominance of the drift field. Remarkably, this transport regime enables device operation at picowatt power levels while sustaining the on/off ratio exceeding $10^5$. We also compared HC-FET under parallel and anti-parallel field configurations using dual side-gates positioned symmetrically around the channel (fig. S18). The parallel configuration yielded significantly higher current and mobility, reconfirming that in-plane gate field mediated carrier acceleration.

Figure 2E maps the electron mobilities against drain voltage and gate bias at room temperature for both $MoS_2$ HC-FET and BG-FET. Notably, in HC-FET, there is a significant dependence of mobility on $V_{DS}$. At low $V_{DS}$, mobility is high but decreases considerably as the drain voltage increases. This trend differs with conventional FETs, where the mobility, according to the standard mobility equation, is normalized by $V_{DS}$, implying that field-effect-mobility mostly remains independent with variations in drain voltage. In HC-FET, the reduction in mobility at high $V_{DS}$ results mainly from carrier-carrier scatterings.

The temperature dependent mobility behavior in HC-FET (Fig. 3A, fig. S19) is in sharp contrast to the phonon- and impurity-limited mobility seen in conventional BG-FETs (Fig. 3B). The high-carrier mobility at elevated temperatures justifies the field-driven hot-carrier mechanism by surpassing phonon scatterings . At room temperature, reduced phonon and impurity scatterings in HC-FET are further substantiated by Raman and Photoluminescence (PL) measurements of $MoS_2$. The observed downshifts (1.44 $cm^{-1}$) of Raman active $A_{1g}$ mode corresponds to electron accumulations in $MoS_2$ from the PMN-PT substrate(*31*), while the full-width at half-maximum (FWHM) for $A_{1g}$ mode was significantly reduced (fig. S20). Since the linewidth is proportional to the electron-phonon coupling strength ($\overline{g^*}$), the linewidth reduction indicates weakening of $\overline{g^*}$ by quenching the available phonon modes for energy dissipation on PMN-PT substrate despite accumulated electrons(*31*). The $\overline{g^*}$ has inverse-square relation to the mobility ($\mu \propto 1/\ \overline{g^{*2}}$)(*32*). Therefore, a slight decrease in $\overline{g^*}$ could cause a greater effect in mobility improvement.

From a slight up-shift (0.03 $cm^{-1}$) of the $E^1_{2g}$ Raman mode on PMN-PT substrate relative to $SiO_2$ substrate, we also estimated the tensile strain of 0.05% (Supplementary materials section 2.5), much smaller than 1% required to enhance substantial strain-induced mobility(*8*, *33*). The strain mapping from FEM simulation (Supplementary materials section 2.6 and fig. 21) further rules out the possibility of strain-induced mobility improvement due to the inverse-piezoelectric effect associated with the PMN-PT substrate. Therefore, the strain plays no significant role in the observed mobility enhancement in HC-FETs.

The PL spectra (fig. S22) of $MoS_2$ on PMN-PT sample exhibit strong quenching, with the intensity reduced by a factor of ~22 with respect to the $MoS_2$ on $SiO_2$ substrate. This intensity suppression arises as most of the carriers are involved in screening process rather than contributing to radiative recombination. Nevertheless, reasonable signature of neutral exciton peaks ($X^0_A$ and $X^0_B$) remain visible, with no detectable trion contribution. The observed neutral exciton peaks also appear redshifted relative to the $SiO_2$-supported sample, which can be attributed to the combined effect of strong dielectric screening from the high-$\kappa$ substrate, ferroelectric polarization field, and subtle strain transfer. In contrast, dominant A-trion ($X^-_A$) state was observed on $SiO_2$ substrate, consistent with the literature(*34*). This is rather surprising considering the accumulation of electrons in $MoS_2$ from PMN-PT substrate. The absence of trion peak can be interpreted as a result of efficient screening of charge impurities on PMN-PT substrate. The charge carriers would otherwise form trions, become localized around the defect centers and thus no longer participated in trion formation. To derive further insights of screening, the frequency and bias dependent metal-insulator-semiconductor capacitance was measured for $MoS_2$/PMN-PT HC-FET (fig. S23). Conventional high-low frequency method(*35*) was used to extract the trap density ($D_{it}$) (Supplementary materials section 2.7). On PMN-PT, $D_{it}$ (corresponding to all the trap states) is lower by two orders of magnitude than that of the $SiO_2$/$MoS_2$ sample, reconfirming efficient screening of charge impurities on PMN-PT substrate.

**Emergence of interaction-driven correlation**

As Coulomb impurities are efficiently screened, the HC-FET is expected to exhibit even higher mobility at low temperature due to the frozen lattice vibrations. However, our experimental observations show a paradoxical reduction in mobility. The temperature-dependent transfer characteristics (Fig. 3C) reveal that at low temperatures, a series of quasi-periodic, gate-tunable fluctuations emerge atop an otherwise conventional FET curve. These features become progressively damped with increasing temperature and carrier density. At low temperature and low carrier density, the suppressed background disorder allows electron-electron repulsion to produce a quasi-periodic correlated potential landscape that guides carrier transport in $MoS_2$ channel. However, the fluctuating conduction may also have different origins like Coulomb blockade or phase coherence.

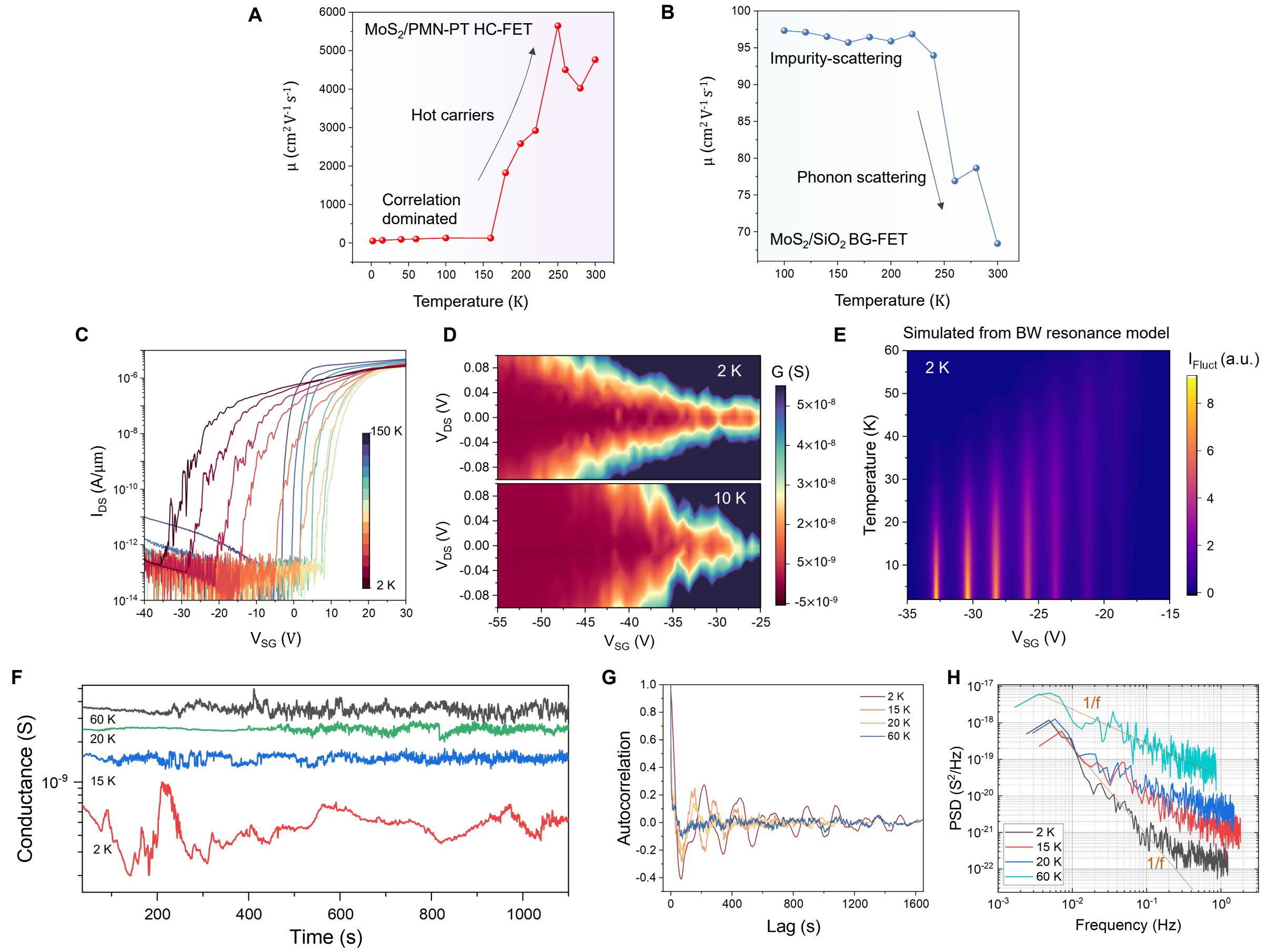

**Fig. 3| Emergence of quantum correlation at low temperatures**. (**A-B**) Mobility of $MoS_2$ HC-FET (**A**) and BG-FET (**B**) *vs* temperature. (**C**) Temperature-dependent transfer characteristics of HC-FET at $V_{DS}$ = 1V, revealing clear transition from quantum fluctuation to lateral field-driven transport. (**D**) Quasi-periodic fluctuation patterns in conductance mapping at 2 K and 10 K. (**E**) Simulated fluctuation features from coupled Fermi-Thomas-Poisson equation. (**F**) Temporal fluctuations at different temperatures. (**G-H**) Autocorrelation functions (**G**) and power spectral densities (**H**) corresponding to the temporal fluctuation data at (**F**).

To gain deeper insight, we plot 2D conductance mapping (Fig. 3D) where the quasi-oscillatory features become significantly weakened at higher $V_{DS}$ at 2K. This feature is further smeared thermally at 10 K. Notably, the absence of any regular Coulomb diamond structure confirms that the fluctuations do not arise from Coulomb blockade, consistent with the lack of lithographically defined confinement in our system(*36*). Instead, the suppression of fluctuations at higher drain bias is indicative of increased electron kinetic energy, disrupting the interaction-induced spatial order. To disentangle the fluctuation features from the semi-classical field-effect behavior, we subtract a smooth FET background from the measured transfer curves, revealing reproducible quasi-periodic peaks across different temperatures (fig. S24 a-e). We simulate this behavior using a phenomenological resonance model derived from a self-consistent Thomas-Fermi-Poisson equation that incorporates electron-electron interactions, ferroelectric screening and spatial potential modulation (Fig. 3E, fig. S24 f, Supplementary materials section 2.8). Notably, the simulated potential profile exhibits quasi-periodic valleys capable of hosting localized resonant states (fig. S25), providing a direct microscopic basis for the reproducible fluctuation features seen in transport measurements at low temperatures. The resonant tunneling features are well described by the Breit-Wigner (BW) form modulated by an electrostatic envelope function associated to the spatial gate coupling strength and density dependent screening effects.

We further examined temporal dynamics of the fluctuations (Fig. 3F-G, fig. S26 a-c). At 2 K, slow non-stationary fluctuation appears with long autocorrelation that gradually dissolves as the temperature increases (fig. S27). Power spectral density (PSD) analysis reveals strong deviation from standard telegraph noise or flicker noise with a transition to suppressed $1/f$ noise at higher temperatures (Fig. 3H, fig. S26 d). Magnetic field-dependent suppression of fluctuations around deep-subthreshold regions was also observed, attributing to the carrier delocalization (fig. S28). However, the fluctuation amplitude does not scale with the conductance quantum ($e^2/h$) and the device dimensions are well beyond the phase coherent length at 2 K(*37*). This reinforces the interpretation that interaction-induced spatial modulation rather than extrinsic disorder, random telegraph noise or phase coherent interference underlies the observed phenomena. Upon increasing temperature, the system undergoes transition from correlation-limited transport to one governed by delocalized hot-carrier dynamics. Evidently, this crossover cannot be accompanied by a sharp phase transition, but by a progressive melting of the ordered landscape.

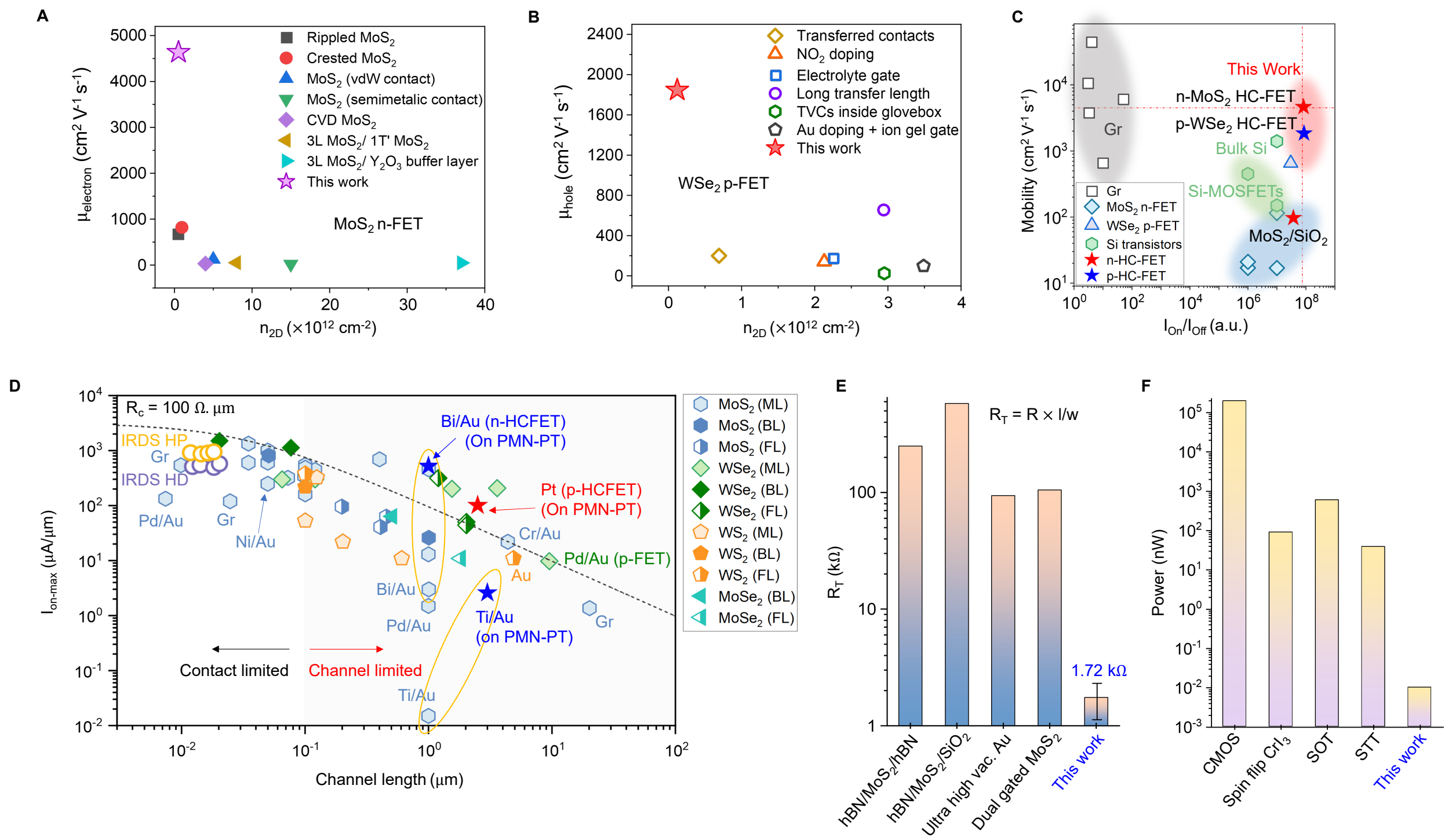

**Fig. 4| Benchmarking HC-FETs against the state-of-the-art 2D transistors.** (**A**) Electron mobility *vs* carrier density in monolayer $MoS_2$ HC-FET with the state-of-the-art $MoS_2$ FETs. (**B**) Hole mobility *vs* carrier density in monolayer $WSe_2$ p-type HC-FET with other $WSe_2$ p-type FETs. (**C**) Mobility *vs* on/off ratio in HC-FETs with other materials including Gr, TMDs and Si-transistors. (**D**) Maximum on-current *vs* channel length of n- and p-type HC-FETs with the state-of-the-art 2D-transistors. (**E-F)** Total resistance and static power dissipation in HC-FET with respect to other high-performance devices.

Figure 4A, B summarizes mobility of the state-of-the-art FETs. Both electron and hole mobilities in HC-FETs for monolayer TMDs exceed the previous records even including multilayer TMDs(*7*, *11*)·(*38–47*). Figure 4C depicts the mobility versus on/off current ratio, illustrating that HC-FETs maintain the high on/off ratio characteristic of TMDs, while achieving graphene-like mobility, significantly outperforming traditional materials(*48*). Figure 4D compares the maximum on-current of our HC-FETs, obtained at a drain-source voltage $V_{DS}$ = 1 V and maximum gate bias, with those reported for other 2D FETs. Notably in long-channel limit, where the current is generally mobility limited(*49*), our HC-FETs exhibited significantly improved $I_{on}$ even for Schottky type (Ti/Au) contacts (fig. S29). Therefore, the improvement is not limited to a specific contact metal. The comprehensive transport performance of HC-FETs is summarized in table S3. In all key FET performance matrices HC-FET offers better performance compared to other state-of-the-art technologies. Figures 4E,F further benchmark the total resistance(*50*, *51*) and static power consumption(*52–54*), respectively, confirming remarkably low resistance and record-low power dissipation. Unlike nonvolatile spintronic memory, which eliminate static power at the cost of write energy and speed, our HC-FETs offer sub-nanowatt active power dissipation during logic operation, suggesting competitive efficiency for volatile logic at room temperature.

**Conclusion**

The combination of thermal fragility, quasi-order, and spatial correlation places the HC-FET system in a regime proximate to soft, glassy phases known from strongly interacting electron systems. We emphasize that this behavior is not indicative of a fully ordered Wigner crystal or a disorder driven Coulomb glass, rather conform with a correlation-stabilized disordered state, where long-range collective behavior emerges from many-body interactions in a flattened extrinsic disorder background. Enabled by dielectric engineering, this intermediate phase bridges the gap between conventional diffusive transport and correlated localization, offering a new experimental way to probe mesoscopic order in TMD monolayers. Consequently, we demonstrate that the same HC-FET architecture achieves record-high electron and hole mobilities in monolayer TMDs, while preserving high on/off ratio, steep SS, and enhanced on-current density at room temperature. These advances stem from the synergistic effects of in-plane ferroelectric polarization and strong dielectric screening, which help to suppress scattering and enable efficient hot-carrier transport. While optimization of gate geometry and scalable synthesis remain important future steps, this study presents a rare convergence of emergent many-body phenomena and high-performance in 2D semiconductor devices.

## References

1. Qiu, H. *et al.* Hopping transport through defect-induced localized states in molybdenum disulphide. *Nat. Commun.* **4**, 3642 (2013).
2. Ghatak, S., Pal, A. N. & Ghosh, A. Nature of electronic states in atomically thin $MoS_2$ field-effect transistors. *ACS Nano* **5**, 7707–7712 (2011).
3. Xiang, Z. *et al.* Imaging quantum melting in a disordered 2D Wigner solid. *Science.* **388**, 736–740 (2025).
4. Lee, D. H., Choi, S. J., Kim, H., Kim, Y. S. & Jung, S. Direct probing of phonon mode specific electron–phonon scatterings in two-dimensional semiconductor transition metal dichalcogenides. *Nat. Commun.* **12**, 4520 (2021).
5. Cheng, L. & Liu, Y. What Limits the intrinsic mobility of electrons and holes in two dimensional metal dichalcogenides? *J. Am. Chem. Soc.* **140**, 17895–17900 (2018).
6. Cui, X. *et al.* Multi-terminal transport measurements of $MoS_2$ using a van der Waals heterostructure device platform. *Nat. Nanotechnol.* **10**, 534–540 (2015).
7. Shen, P. C. *et al.* Ultralow contact resistance between semimetal and monolayer semiconductors. *Nature* **593**, 211–217 (2021).
8. Hosseini, M., Elahi, M., Pourfath, M. & Esseni, D. Strain-induced modulation of electron mobility in single-layer transition metal dichalcogenides $MX_2$ ($M$ = Mo, W; $X$ = S, Se). *IEEE Trans. Electron Devices* **62**, 3192–3198 (2015).
9. Chen, Y. *et al.* Mobility enhancement of strained $MoS_2$ transistor on flat substrate. *ACS Nano* **17**, 14954–14962 (2023).
10. Datye, I. M. *et al.* Strain-enhanced mobility of monolayer $MoS_2$. *Nano Lett.* **22**, 8052–8059 (2022).
11. Liu, T. *et al.* Crested two-dimensional transistors. *Nat. Nanotechnol.* **14**, 223–226 (2019).
12. Kayal, A. *et al.* Mobility enhancement in CVD-grown monolayer $MoS_2$ via patterned substrate-induced nonuniform straining. *Nano Lett.* **23**, 6629–6636 (2023).
13. Jena, D. & Konar, A. Enhancement of carrier mobility in semiconductor nanostructures by dielectric engineering. *Phys. Rev. Lett.* **98**, 136805 (2007).
14. Bolshakov, P. *et al.* Improvement in top-gate $MoS_2$ transistor performance due to high quality backside $Al_2O_3$ layer. *Appl. Phys. Lett.* **111**, 032110 (2017).
15. Ong, Z. Y. & Fischetti, M. V. Mobility enhancement and temperature dependence in top-gated single-layer $MoS_2$. *Phys. Rev. B* **88**, 165316 (2013).
16. Simon M. Sze, K. K. Ng. *Physics of Semiconductor Devices*. (Wiely, ed. 3, 2007).
17. Schöffmann, P. *et al.* Strain and charge contributions to the magnetoelectric coupling in $Fe_3O_4$/PMN-PT artificial multiferroic heterostructures. *New J. Phys.* **24**, 123036 (2022).
18. Shvartsman, V. V. & Kholkin, A. L. Domain structure of $0.8Pb(Mg_{1/3}Nb_{2/3})O_3$ - $0.2PbTiO_3$ studied by piezoresponse force microscopy. *Phys. Rev.* **69**, 014102 (2004).
19. Bian, J. *et al.* Fingerprints of relaxor ferroelectrics: Characteristic hierarchical domain

configurations and quantitative performances. *Appl. Mater. Today* **21**, 100789 (2020).

20. Frey, I., Constant, E., Boittiaux, B., Hess, K. & Hess, K. Novel real-space hot-electron transfer devices. *IEEE Electron Device Letters* **4**, 334–336 (1983).

21. Liu, C. *et al.* A hot-emitter transistor based on stimulated emission of heated carriers. *Nature* **632**, 782–787 (2024).

22. Blatter, G. & Baeriswyl, D. High-field transport phenomenology: Hot-electron generation at semiconductor interfaces. *Phys. Rev. B* **36**, 6446–6464 (1987).

23. Ma, N. & Jena, D. Charge scattering and mobility in atomically thin semiconductors. *Phys. Rev. X* **4**, 014102 (2014).

24. Spivak, B., Kravchenko, S. V., Kivelson, S. A. & Gao, X. P. A. Colloquium: Transport in strongly correlated two dimensional electron fluids. *Rev. Mod. Phys.* **82**, 1743–1766 (2010).

25. Sebastian, A., Pendurthi, R., Choudhury, T. H., Redwing, J. M. & Das, S. Benchmarking monolayer $MoS_2$ and $WS_2$ field-effect transistors. *Nat. Commun.* **12**, 693 (2021).

26. Lombardi, C., Manzini, S., Saporito, A. & Vanzi, M. A physically based mobility model for numerical simulation of nonplanar devices. *IEEE Trans. Comput. Des. Integr. Circuits Syst.* **7**, 1164–1171 (1988).

27. Canali, C., Minder, R. & Ottaviani, G. Electron and hole drift velocity measurements in silicon and their empirical relation to electric field and temperature. *IEEE Trans. Electron Devices* **22**, 1045–1047 (1975).

28. Lee, Y. H. Approaching the quantum limit of contact resistance in van der Waals layered semiconductors. *Science.* **384**, 6702 (2024).

29. Salahuddin, S. & Datta, S. Use of negative capacitance to provide voltage amplification for low power nanoscale devices. *Nano Lett.* **8**, 405–410 (2008).

30. Xu, L. *et al.* Ferroelectric-modulated $MoS_2$ field-effect transistors as multilevel nonvolatile memory. *ACS Appl. Mater. Interfaces* **12**, 44902–44911 (2020).

31. Chakraborty, B. *et al.* Symmetry-dependent phonon renormalization in monolayer $MoS_2$ transistor. *Phys. Rev. B - Condens. Matter Mater. Phys.* **85**, 161403 (2012).

32. Cheng, L., Zhang, C. & Liu, Y. Why Two-Dimensional Semiconductors Generally Have Low Electron Mobility. *Phys. Rev. Lett.* **125**, 177701 (2020).

33. Velický, M. *et al.* Strain and charge doping fingerprints of the strong interaction between monolayer $MoS_2$ and gold. *J. Phys. Chem. Lett.* **11**, 6112–6118 (2020).

34. Pandey, J. & Soni, A. Unraveling biexciton and excitonic excited states from defect bound states in monolayer $MoS_2$. *Appl. Surf. Sci.* **463**, 52–57 (2019).

35. Zhao, P. *et al.* Evaluation of border traps and interface traps in $HfO_2/MoS_2$ gate stacks by capacitance-voltage analysis. *2D Mater.* **5**, 031002, (2018).

36. Kotekar-Patil, D., Deng, J., Wong, S. L. & Goh, K. E. J. Coulomb blockade in etched single- and few-layer $MoS_2$ nanoribbons. *ACS Appl. Electron. Mater.* **1**, 2202–2207 (2019).

37. Matsuo, S. *et al.* Experimental proof of universal conductance fluctuation in quasi-one-dimensional epitaxial $Bi_2Se_3$ wires. *Phys. Rev. B* **88**, 155438 (2013).

38. Ng, H. K. *et al.* Improving carrier mobility in two-dimensional semiconductors with rippled materials. *Nat. Electron.* **5**, 489–496 (2022).

39. Wang, Y. *et al.* Van der Waals contacts between three-dimensional metals and two-dimensional semiconductors. *Nature* **568**, 70–74 (2019).

40. Kang, K. *et al.* High-mobility three-atom-thick semiconducting films with wafer-scale homogeneity. *Nature* **520**, 656–660 (2015).

41. Kappera, R. *et al.* Phase-engineered low-resistance contacts for ultrathin $MoS_2$ transistors. *Nat. Mater.* **13**, 1128–1134 (2014).

42. Zou, X. *et al.* Interface Engineering for high-performance top-gated $MoS_2$ field-effect transistors. *Adv. Mater.* **26**, 6255–6261 (2014).

43. Wu, Z. *et al.* Defects as a factor limiting carrier mobility in $WSe_2$: A spectroscopic investigation. *Nano Res.* **9**, 3622–3631 (2016).

44. Fang, H. *et al.* High-performance single layered $WSe_2$ p-FETs with chemically doped contacts. *Nano Lett.* **12**, 3788–3792 (2012).

45. Liu, Y. *et al.* Low-resistance metal contacts to encapsulated semiconductor monolayers with long transfer length. *Nat. Electron.* **5**, 579-585 (2022).

46. Jung, Y. *et al.* Transferred via contacts as a platform for ideal two-dimensional transistors. *Nat. Electron.* **2**, 187–194 (2019).

47. Chen, C. H. *et al.* Hole mobility enhancement and p-doping in monolayer $WSe_2$ by gold decoration. *2D Mater.* **1**, 034001 (2014).

48. Jin, Y. *et al.* Coulomb drag transistor using a graphene and $MoS_2$ heterostructure. *Commun. Phys.* **3**, 189 (2020).

49. Zhuo, F. *et al.* Modifying the Power and performance of 2-dimensional $MoS_2$ field effect transistors. *Research* **6**, 0057 (2023).

50. Mondal, A. *et al.* Low ohmic contact resistance and high on/off ratio in transition metal dichalcogenides field-effect transistors via residue-free transfer. *Nat. Nanotechnol.* **19**, 34–43 (2024).

51. English, C. D., Shine, G., Dorgan, V. E., Saraswat, K. C. & Pop, E. Improved contacts to $MoS_2$ transistors by ultra-high vacuum metal deposition. *Nano Lett.* **16**, 3824–3830 (2016).

52. Fu, Z. E. *et al.* Tunneling current-controlled spin states in few-layer van der Waals magnets. *Nat. Commun.* **15**, 3630 (2024).

53. Chen, S. *et al.* Magnetic Switching in Monolayer 2d diluted magnetic semiconductors via spin-to-spin conversion. *Adv. Funct. Mater.* **35**. 2418647 (2025).

54. Wang, M. *et al.* Current-induced magnetization switching in atom-thick tungsten engineered perpendicular magnetic tunnel junctions with large tunnel magnetoresistance. *Nat. Commun.* **9**, 671 (2018).

**Data and materials availability:** All data needed to support the conclusions of this article are available in the main text or the supplementary materials.

**Acknowledgements**

M.-K.J. acknowledges the support from National Research Foundation of Korea (NRF) grant funded by the Korea government (MSIT) (RS-2025-00514053, RS-2023-00254934).

**Author contributions**

D.D. and H.B. initiated the project under the supervision of Y.H.L.. D.D. fabricated all the devices and performed transport measurements and optical spectroscopies. H.B. contributed to CAD design with D.D.. D.D measured KPFM with the help of S.-G.L. and W.C.. K.K.K. supplied CVT crystals. M.-K.J. participated in PSD data analysis. All the simulations and data processing were carried out by D.D. with the input from Y.H.L.. D.D. and Y.H.L. analyzed all the data and prepared the manuscript. Y.W. and M.C. contributed to data interpretation. All the authors verified the manuscript. Y.H.L. supervised the entire work.

**Competing interests**

The authors declare no competing interests.

**Additional information**

**Supplementary information**

The online version contains supplementary material available at

## 1. Materials and Methodes

### 1.1 Device fabrication

$MoS_2$ and $WSe_2$ monolayers were mechanically exfoliated from synthetic bulk crystals (2D Semiconductors, USA or lab-grown CVT single crystals) on top of a poly(vinyl alcohol) (PVA, Sigma Aldrich)-coated $SiO_2$/Si substrate, followed by spin-coating a PMMA A4 or Poly-propylene carbonate (PPC, Sigma Aldrich) layer. The monolayer flakes were primarily identified by the optical contrast and further verified by atomic force microscopy (AFM) or Raman spectroscopy. The exfoliated flakes were subsequently transferred using a conventional wet transfer method onto the commercial PMN-PT (001) substrate of a thickness 0.3 mm. Electrodes were patterned using standard e-beam lithography process after spin coating (500 rpm, 5 s and 3000 rpm, 60 s) 495 PMMA A4/ 950 PPMA A4 double layer, followed by post baking at 120° C for 2 minutes, and spin coating of e-beam anti-charging agent (DisChem). The anti-charging agent was removed by soaking in IPA for 20 s prior to following standard PMMA developing steps. For $SiO_2$ BG-FETs, a similar process was followed to maintain identical sample conditions.

20/50 nm of Bi/Au (for n-type contacts) or 25 nm of Pt (for p-type contacts) metal electrodes were deposited (deposition rate was optimized to avoid 2D channel/metal interface degradation) using an ultra-high vacuum thermal evaporator installed inside an Ar-filled glove box or using RF sputtering system (deposition pressure ~$10^{-7}$ mTorr). Metal lift-off was done with acetone followed by cleaning in IPA and ethanol respectively.

### 1.2 Electrical transport measurements

The FET performance of all the devices was characterized at the vacuum probe station (~$10^{-6}$ torr) using Keithley 4200A semiconductor parameter analyzer equipped with remote preamplifier modules (0.1 fA resolution). Temperature-dependent $I_{DS}$-$V_{DS}$ were measured using a Dynacool physical property measurement system (Quantum design). Ferroelectric polarization-drive voltage hysteresis curves were measured by applying triangular voltage pulses (10 kHz) using a Radiant RT66C multiferroic tester.

Capacitance measurements were conducted using Keysight B1500A equipped with N131A CMU module. Proper calibration was done to rule out parasitic capacitance contribution from the cables and probe tips. The measured capacitance was further verified with HP 4285A LCR meter.

**Extraction of mobility and carrier density**.

The carrier density was estimated from the $n = C_{\text{in}}(V_{GS} - V_{th})/e$. Two-probe field-effect ($\mu$) mobility was extracted from the linear region of the transfer characteristics using the expression, $\mu = \frac{\partial I_{DS}}{\partial V_{GS}}\left(\frac{L}{W C_{\text{in}} V_{DS}}\right)$, where $L/W$ is the length to width ratio of the channel and $C_{\text{in}}$ is the gate capacitance in F/cm$^2$. For the side-gate on PMN-PT substrate, capacitance was experimentally determined by aforementioned *C*-*V* measurement technique. In some of the devices, the gate over-drive voltage was, $|V_{GS} - V_{th}| < V_{DS}$, where the saturation mobility is extracted by using $\mu = \left( \frac{\partial \sqrt{|I_{DS}|}}{\partial V_{GS}} \right)^2 \left( \frac{2L}{W C_{\text{in}}} \right)$.

Due to the formation of ohmic contact, contact resistance has no significant contribution in determining two-terminal mobility. However, to ensure the accuracy of the estimated mobility value, four-terminal mobility was simultaneously measured and calculated following the expression, $\mu = \frac{\partial (I_{DS}/V_{12})}{\partial V_{GS}} \left( \frac{L_{12}}{W C_{\text{in}}} \right)$, where $V_{12}$ is the measured voltage between the inner probes separate by $L_{12}$.

## 1.3 Optical spectroscopy

The Stokes lines of the Raman signal were collected with 1800 lines/mm grating at WITec alpha 300 with 532 laser excitations. The spectrum was calibrated by the 520 cm$^{-1}$ Si characteristic peak prior to the measurement. Photoluminescence was carried out at NT-MDT with 532 nm laser excitation (via Cobalt 05-01 series laser source) under ambient conditions. The spectrum could be quenched easily on PMN-PT substrate. Therefore, a relatively higher excitation power of 0.6 mW was used to improve signal to noise ratio.

## 2. Supplementary Text

### 2.1 Simulation for dielectric screening

When a static charge is placed inside the 2D semiconductor channel, the original Coulomb potential of defects is attenuated due to the formation of an opposite polarity charge-cloud around it (fig. S1 a,b). The Coulomb potential can be screened within the bulk semiconductors, whereas the unscreened potential extends beyond the channel in atomically thin TMDs and therefore influenced by the surrounding dielectric environment. For simplicity, we consider here a system of monolayer $MoS_2$ ($\varepsilon_2$= 6.4) sandwiched between two dielectric environments ($\varepsilon_1$ and $\varepsilon_3$) and a static point charge-impurity located at $(\rho_0,\ z_0=0)$.

The unscreened Coulomb potential is calculated (fig. S1 c-e), following the method of image charges(*1*, *2*). The impurity charge located inside $MoS_2$ results in the formation of an infinite series of point image charges at $z=na$ where $n=\ \pm1,\ \pm2,$....The amplitude of the $n^{\text{th}}$ image charge is given by $e\gamma_{\pm}^{|n|}$, where the dielectric mismatch factors are defined as $\gamma_{+}=(\varepsilon_2-\varepsilon_1)/(\varepsilon_2+\varepsilon_1)$ and $\gamma_{-}=(\varepsilon_2-\varepsilon_3)/(\varepsilon_2+\varepsilon_3)$, corresponding to the upper and lower dielectric media, respectively. The net unscreened potential experienced by the free carriers is given by $V(\vec{\rho},z)=\sum_{n=-\infty}^{\infty}\frac{e\gamma_{\pm}^{|n|}}{4\pi\varepsilon_0\varepsilon_2\sqrt{|\rho-\rho_0|^2+|z-z_n|^2}}$, where $e$ is the electronic charge, $\vec{\rho}$ is the in-plane position vector of electron relative to charge impurity and $\varepsilon_0$ is the free space permittivity(*3*). It is evident that when the relative permittivity of the surrounding media is lower than that of $MoS_2$ ($\varepsilon_1$and $\varepsilon_3<\varepsilon_2$), all image charges have the same sign, leading to larger effective Coulomb potential (fig. S1 c, d). In contrast, for PMN-PT due to its higher dielectric constant, the magnitude of image charges alternates between positive and negative values depending on $n$, resulting in the lower Coulomb potential seen by the free carriers (fig. S1 e).

### 2.2 Simulation for polarizability and scattering potential

The Poisson's equation for a static charge impurity in monolayer $MoS_2$ is given by(*4*),

$$-\nabla^2 \phi_{Scr}(\vec{q}, z) = \frac{\rho_{imp} + \rho_{scr}}{\varepsilon_0} \tag{2.1}$$

Where, $\rho_{imp}$ and $\rho_{scr}$ are the impurity charge and the induced charges that cause scattering.

The in-plane screened scattering potential can be written as:

$$\phi_{Scr}(\vec{q}, z) = \frac{e^2 G_q(0,0)}{\epsilon_{2D}} \tag{2.2}$$

Where $\vec{q}$ and $G_q(0,0)$ are wavevector and Green's function solution of Poisson's equation, respectively. For a system consisting of Air/$MoS_2$/PMN-PT $G_q(0,0)$ is given as(*5*), $G_q(0,0) = \frac{1}{\varepsilon_0(1+\varepsilon_{PMN-PT})q}$ *(2.3)*

At finite temperature, the generalized static screening function by the free carriers is given by,

$$\epsilon_{2D} = 1 + e^2 G_q(0,0)\Pi(\vec{q}, T,\ E_F), \tag{2.4}$$

which exclusively depends on the temperature-dependent static polarizability function of the mobile carriers,

$$\Pi(\vec{q}, T,\ E_F) = \int_0^\infty d\mu \frac{\Pi(\vec{q}, 0,\ \mu)}{4K_B T\ \cosh^2(E_F - \mu/2K_B T)}. \tag{2.5}$$

Here $\mu$ is the chemical potential, not refer to carrier mobility. The polarizability function at zero temperature is given by,

$\Pi(\vec{q}, 0,\ \mu) = -\frac{gm^*}{2\pi\hbar^2}[1-\Theta(q-2k_F)\sqrt{1-(\frac{2k_F}{q})^2}]$, where $g(=4)$ is the spin and valley degeneracy, $m^*(=0.48m_0)$ is the effective mass of electron and the Fermi wavevector, $k_F = \sqrt{2m^*\mu}/\hbar$.

As temperature increases, free carriers try to restore thermal equilibrium that randomizes the carrier momenta, hence weakens the screening. Meanwhile, when a side-gate bias is applied in HC-FET, the in-plane polarization in PMN-PT substrate induces an electric field $E_{ind} = P_{DG}/\varepsilon_0 \chi_e$, which causes a shift in electron distribution in *k*-space, $\Delta k = \frac{e\ E_{ind}\ \tau}{\hbar}$ where, $\tau$ is the momentum relaxation time. The corresponding change in the Fermi energy due to the shift is given by, $\Delta E_F = \frac{\hbar^2 \Delta k (2k_F + \Delta k)}{2m^*}$. Therefore, the modified Fermi energy for $\Pi(\vec{q}, T, E_F)$ is given as, $E_F^* = E_F + \Delta E_F$.

Extended Data Fig. 2 plots the normalized static polarizability as the function of perturbation wavevector at different temperatures. The shift in Fermi surface by the externally induced field in HC-FET clearly leads to higher polarizability at certain $\vec{q}$ and temperature, therefore lowering scattering potential.

The intrinsic mobility is given by, $\mu_e = \sigma / en$, where $n$ is the carrier density of 2D channel and $\sigma$ is the conductivity.

For single layer $MoS_2$, using the generalized expression of $\sigma$, the mobility can be rewritten as,

$$\mu_e = \frac{e}{\pi n \hbar^2 K_B T} \int f(E)[1 - f(E)] \tau(E_q) E dE, \qquad (2.6)$$

where $f(E)$ is the equilibrium Femi-Dirac distribution function and the scattering rate (*i.e.*, inverse of the relaxation time) is given by,

$$\frac{1}{\tau(E_q)} = \frac{n_{imp}}{2\pi \hbar} \int dq' |\phi_{Scr}|^2 (1 - \cos\theta) \delta(E_q - E_{q'}) \qquad (2.7)$$

The $n_{imp}$ is the impurity density in $MoS_2$ and $\theta$ is the scattering angle between $q'$ and $q$. The relation between $\tau(E_q)$ and mobility clearly indicates that weakening the scattering potentially led to enhanced mobility in HC-FETs.

### 2.3 Kelvin-probe force microscopy (KPFM)

KPFM measurement was performed using a Hitachi E-sweep AFM system, with a drive voltage of 1 V applied through the rhodium (Rh)-coated tip of the Si cantilever. Conventional KPFM chip setup is customized (fig. S6 a) to apply drain and gate bias externally by dual channel Keithley 2636B source measure unit. As polymer residues on the FET-channel surface can severely affect the KPFM signal, special care was taken during sample preparation to minimize the fabrication-process induced contamination. The FET was isolated from the metallic sample holder, while the source terminal was connected to common floating ground. When the tip is electrostatically connected (by approaching to a close proximity) to the sample, the Fermi level was lined up to shift the vacuum level by the amount of the work function (WF) difference between sample and tip. Due to the capacitive coupling, the tip experiences a net electric field. In KPFM, an additional DC tip bias ($V_{CPD}$) along with an AC bias (of 27.11 kHz) is applied to compensate the work function difference that renormalizes the vacuum level and electrostatic force experienced by the tip(*6*). Therefore: $WF_{MoS_2} = WF_{Rh} - eV_{CPD}$. In conventional FET, the application of gate bias results in vertical band modulation in the channel, while the application of $V_{DS}$ induces lateral band bending(*7*). In contrast, side-gate bias in HC-FET leads to significant lateral band modulation, accompanied by nominal vertical band modulation. The relative contact potential difference (CPD) along the $MoS_2$ channel corresponds to the non-local gate-field induced enhancement in lateral band bending that promotes the primary drift-field caused by the drain bias (fig. S6 d). Conversely, the off-state in HC-FET arises from the chemical potential lowering by the vertical component of the gate field as well as the competing in-plane gate-field that suppresses the majority carrier flow by disfavoring the channel-band slope. The band modulation through the side-gate offers a distinct advantage over traditional vertical-gate designs, where such dynamic control is not possible.

### 2.4 <u>Finite element method (FEM) of mobility simulation</u>

FEM simulation is performed to understand the effect of change in gate geometry as well as varying dielectric environments on the carrier mobility. The given input reference mobility is corrected for acoustic phonon and surface roughness scattering based on the Lombardi surface mobility model and further corrected for optical phonon and high field velocity saturation using the Caughey-Thomas model. The reference mobilities are obtained from the experimental reports which are used as the input for the mobility correction models.

The mathematical description of the Lombardi model is as follows(*8*);

$\frac{1}{\mu_{n,Lo}} = \frac{1}{\mu_{¿,n}} + \frac{1}{\mu_{ac,n}} + \frac{1}{\mu_{sr,n}}$, where $\mu_{¿,n}$ is the input reference, $\mu_{ac,n}$ is the acoustic phonon limited term and $\mu_{sr,n}$ is the scattering due to surface roughness. From the equation it could clearly be understood that the resultant mobility must be smaller than the smallest constituent mobility value. Both the $\mu_{ac,n}$ and $\mu_{sr,n}$ show the inverse dependence on the perpendicular field strength as follows;

$$\mu_{ac,n} = \frac{\mu_{1,n}}{\left(E_{\perp,n}/E_{Ref}\right)} + \frac{\mu_{2,n}\left(N/N_{Ref}\right)^{\beta_n}}{\left(E_{\perp,n}/E_{Ref}\right)^{1/3}\left(T/T_{Ref}\right)} \text{ and } \mu_{sr,n} = \frac{\delta_n}{E_{\perp,n}{}^2}.$$

Therefore, lowering the perpendicular electric field may lead to reduced surface scattering, originating from the acoustic phonons and surface roughness.

For $MoS_2$ on $SiO_2$ substrate, the maximum reference mobility value is reported as 200 $cm^2V^{-1}s^{-1}$.(*9*) After applying the Lombardi correction, the mobility drops to ~60 $cm^2V^{-1}s^{-1}$ in $SiO_2$ BG-FET, which is significantly lower than the corrected mobility ~110 $cm^2V^{-1}s^{-1}$ in the $SiO_2$ SG-FET. In contrast, for the given experimental mobility of ~2000 $cm^2V^{-1}s^{-1}$ in HC-FET (provided in fig. S12), after applying Lombardi correction, reference mobility is reduced to ~1590 $cm^2V^{-1}s^{-1}$. This illustrates a considerably smaller reduction in mobility compared to the mobility degradation in $SiO_2$ devices.

Finally, contribution of optical phonon scattering and velocity saturation at high field is corrected with the Caughey-Thomas model as follows(*10*);

$$\mu_{n,CT} = \frac{\mu_{¿,n}}{\left(1+\left(\mu_{¿,n}F_n/v_{sat,n}\right)^{\alpha_n}\right)^{1/\alpha_n}},$$ where the $\mu_{¿,n}$ is the input mobility taken from Lombardi model and $F_n = E_{\parallel,n}$ is the lateral driving electric field of electrons. In BG-FET, the mobility decreases to ~55 $cm^2V^{-1}s^{-1}$ after applying Caughey-Thomas correction, consistent with experimental mobility in conventional BG-FETs. In contrast, the SG-FET shows minimal change in mobility of ~109 $cm^2V^{-1}s^{-1}$ after correction, which is attributed to the nominal contribution of in-plane field from the side-gate on $SiO_2$ substrate. Meanwhile, the corrected mobility in HC-FET drops to ~1446 $cm^2V^{-1}s^{-1}$ after correction.

In conclusion, the overall reduction in mobility from its given reference is 72% for BG-FET, 45.5% for SG-FET and only 27.7% for HC-FET. This reflects that the scattering effect is less prominent in HC-FET compared to other devices such as BG-FET.

### 2.5 <u>Estimation of strain in HC-FET</u>

As reported in previous literature, a large biaxial tensile strain of over 1% can cause up to 2-3 times enhancement of carrier mobility in $MoS_2$(*11*). Therefore, it is necessary to analyze if there is any significant contribution of the strain in many-fold mobility improvements in HC-FET.

The lateral strain originating from lattice mismatch between $MoS_2$ and PMN-PT substrate was determined from the relative shift in $E_{2g}^{1}$ mode (fig. S20)(*12*). $\delta\omega_M = \omega_M^0 - \omega_M = 2\gamma_M \omega_M^0 \varepsilon$, gives the change in frequency in generic Raman mode, where $\omega_M^0$ and $\omega_M$ are the Raman frequencies of M mode in unstrained and strained conditions(*13*). $\gamma_M$ is the Grüneisen parameter (0.64 - 1.1 for $MoS_2$) of M mode and $\varepsilon$ is the biaxial strain(*14*). Here the change in strain percent is determined relative to the reference $MoS_2/SiO_2$ sample. Considering $\gamma_M$ to be 0.82, the estimated strain is 0.05% compared to that on $SiO_2$ substrate.

## 2.6 Simulation of strain percent due to inverse-piezoelectricity in PMN-PT

PMN-PT is known for its high piezoelectric response having one of the highest strain coefficients. When the side-gate bias is applied in HC-FET, associated electric field may cause mechanical deformation (a phenomenon known as inverse-piezoelectric effect) to the PMN-PT substrate, hence inducing to the $MoS_2$ channel. The generated overall strain is the summation of two strains: (1) volumetric strain (related to the bulk modulus of the material) and (2) deviatoric strain (linked to the elastic modulus). Here we calculate the strains in HC-FET using the strain-charge form (FEM):

$$S = s_E T + d^T E$$

$$D = dT + \varepsilon_0 \varepsilon_{rT} E$$

where $S$, $T$, $D$, and $E$ are the strain, stress, displacement and electric field respectively. $s_E$, $\varepsilon_{rT}$, and $d$ represent material parameters like material compliance, permittivity at constant stress and coupling strength(*15*).

Here the piezoelectric strain coefficients of PMN-PT are obtained from the material datasheet provided by the manufacturer(*16*). The strain distributions for parametric sweep of gate and drain voltages are provided in fig. S21. Even at high gate bias ($V_{SG}$ = 10 V), both strain contributions remained $\ll$1%, indicating the improved mobility is definitely not caused by strained $MoS_2$.

## 2.7 Estimation of trap density from *C-V*

Conventional high-low frequency method is employed to extract the trap density from *C-V* data (fig. S23). The trap capacitance is given by(*17*, *18*),

$$C_{\mathrm{it}} = \left( \frac{1}{C_{LF}} - \frac{1}{C_{ins}} \right)^{-1} - \left( \frac{1}{C_{HF}} - \frac{1}{C_{ins}} \right)^{-1}$$

where $C_{LF}$ is the MIS capacitance measured at low frequency (10 kHz) and $C_{HF}$ is the capacitance measured at very high frequency (1 MHz) and $C_{ins}$ is the substrate capacitance. The trap density is estimated from the expression, $D_{\mathrm{it}} = C_{\mathrm{it}} / e$.

## 2.8 Current fluctuations through correlated potential landscape

Charge transport in 2D materials is generally governed by scatterings from electron-phonon, electron-impurity and electron-electron interactions(*19*). At room temperature, phonon and impurity scatterings dominate, effectively masking any contribution from electron-electron interactions. As the temperature decreases, phonon scattering is suppressed due to frozen lattice vibration, leaving impurity scattering as the primary limiting mechanism. In both conventional transport regimes, the role of electron-electron interactions is overshadowed by other dominant scattering channels.

In contrast, in our HC-FET system, the impurities are effectively screened by the PMN-PT substrate. As a result, once phonon scattering weakens with decreasing temperature, electron-electron interactions emerge as key mechanism governing transport in the channel. At sufficiently low carrier density, the kinetic energy of electrons becomes comparable to their mutual Coulomb repulsion. These long-range interactions force electrons to spatially localize across the channel to minimize their interaction energy. This correlated potential landscape effectively acts as induced defects hindering the normal electron transport through the channel.

Theoretically this effect can be formulated starting from the Poisson equation for a charge density $\rho(x,y)$ which induces an electrostatic potential $\phi(x,y)$.

$$\nabla^2 \phi(x,y) = -\frac{\rho(x,y)}{\varepsilon_0 \varepsilon_r} \tag{8.1}$$

Where the charge density can be written as, $\rho(x,y) = -en(x,y)$, and $n(x,y)$ be the density of electrons, $\varepsilon_0$ and $\varepsilon_r$ are free space permittivity and dielectric constant of $MoS_2$ respectively.

Defining potential energy of the electrons, $V(x,y) = -e\phi(x,y)$, Poisson equation *(8.1)* can be rewritten as,

$$\nabla^2 V(x,y) = -\frac{e^2}{\varepsilon_0 \varepsilon_r} n(x,y) \tag{8.2}$$

The potential landscape seen by the electrons depends on the local distribution of electrons. In the degenerate limit, Thomas-Fermi approximation(*20*, *21*) provides the electron density around the local potential energy,

$$n(x,y) = \frac{g_s g_v m^*}{2\pi\hbar^2}\left[E_F - V(x,y)\right].\Theta\left(E_F - V(x,y)\right) \tag{8.3}$$

Where, $m^*$ is the effective mass of electron in $MoS_2$, $g_s$ and $g_v$ refer to spin and valley degeneracy, $\Theta$ is the Heaviside step function to enforce the positive part of the energy and $E_F$ is the local Fermi energy set by the side-gate.

Substituting $n(x,y)$ in equation *(8.2)*, the nonlinear self-consistent Thomas-Fermi-Poisson equation,

$$\nabla^2 V(x,y) + C\,\left[E_F - V(x,y)\right].\Theta\left(E_F - V(x,y)\right) = 0 \tag{8.4}$$

Where $C = \frac{g_s g_v e^2 m^*}{2\pi\hbar^2 \varepsilon_0 \varepsilon_r}$.

The equation *(8.4)* has been solved numerically on a $MoS_2$ grid of dimensions $7\,\mu m \times 2\,\mu m$ and carrier density $10^{-10}$ (around the subthreshold region of the transfer curve) $cm^{-2}$ to demonstrate the emergent spatial correlation in fig. S25. The potential will have the contribution from the electron density, residual charge impurities, ferroelectric dipoles in PMN-PT, and lateral field-induced tilt.

Solutions of the equation *(8.4)* yield a quasi-periodic potential landscape composed of localized valleys at $(x_i, y_i)$, which act as 2D quantum harmonic wells,

$$V(x,y) = V_{0,i} + \frac{1}{2} m^* \omega_i^2 (x - x_i)^2 + \frac{1}{2} m^* \omega_i^2 (y - y_i)^2 \tag{8.5}$$

Where, $V_{0,i}$ is the potential minimum at the $i^{th}$ valley.

Each of these valleys supports discrete bound states $E_i = V_{0,i} + \frac{\hbar}{2}(\omega_{x,i} + \omega_{y,i})$, through which carriers can tunnel when the local chemical potential aligns with $E_i$. The resulting transmission coefficient can be given by the Breit-Wigner resonance formula(*22*, *23*),

$$T(E) = \frac{\Gamma_L \Gamma_R}{(E - E_i)^2 + (\Gamma/2)^2} \tag{8.6}$$

Where $\Gamma_{L,R}$ is the coupling strength into left and right reservoirs and $\Gamma = \Gamma_L + \Gamma_R$ is the total linewidth.

As the shift in Fermi energy is caused by side-gate bias ($V_{SG}$) sweep, mapping gate voltage to Fermi energy, the expression for current is obtained in Lorentzian form(*24*).

$$I_i(V_{SG}, T) = \frac{A_i}{1 + [(V_{SG} - V_{0,i})/\Gamma_i]^2} \tag{8.7}$$

Where is the $A_i$ tunneling amplitude and $\Gamma_i = \Gamma/\alpha$ is the resonance broadening.

In addition to the intrinsic Lorentzian broadening ($\Gamma_i$) due to resonant tunneling, the contribution of each valley is further suppressed by additional electrostatic envelope considering the capacitive coupling to the gate and density-dependent screening (with increasing gate bias, number of carriers increases that enhances the kinetic energy thereby smoothens the induced potential landscape). Here these effects are captured with a broad Lorentzian envelope characterized by a scaling constant $\Lambda$. "$\Lambda$" correlates the gate-voltage range over which screening mechanism suppresses valley localization. The resulting current is thus a product of a sharp quantum tunneling resonance peak the amplitude of which depends on temperature as well as the gate bias.

$$I_i(V_{SG}, T) = \frac{A_i}{1 + [(V_{SG} - V_{0,i})/\Gamma_i]^2} \cdot \frac{1}{1 + [(V_{SG} - V_{0,i})/\Lambda]^2} \tag{8.8}$$

Summing over all valleys reproduces the quasi-periodic current fluctuations (Fig. 3e) experimentally observed in the transfer characteristics,

$$I_i(V_{SG}, T) = \sum_i \frac{A_i}{1 + [(V_{SG} - V_{0,i})/\Gamma_i]^2} \cdot \frac{1}{1 + [(V_{SG} - V_{0,i})/\Lambda]^2} \tag{8.9}$$

We emphasize that the equation *(8.9)* is not intended to describe the full FET transfer characteristic, but rather the reproducible fluctuation peaks that remain after subtracting the drift current baseline. This decomposition reflects the physical separation between semiclassical conductivity and correlation-driven quantum tunneling features.

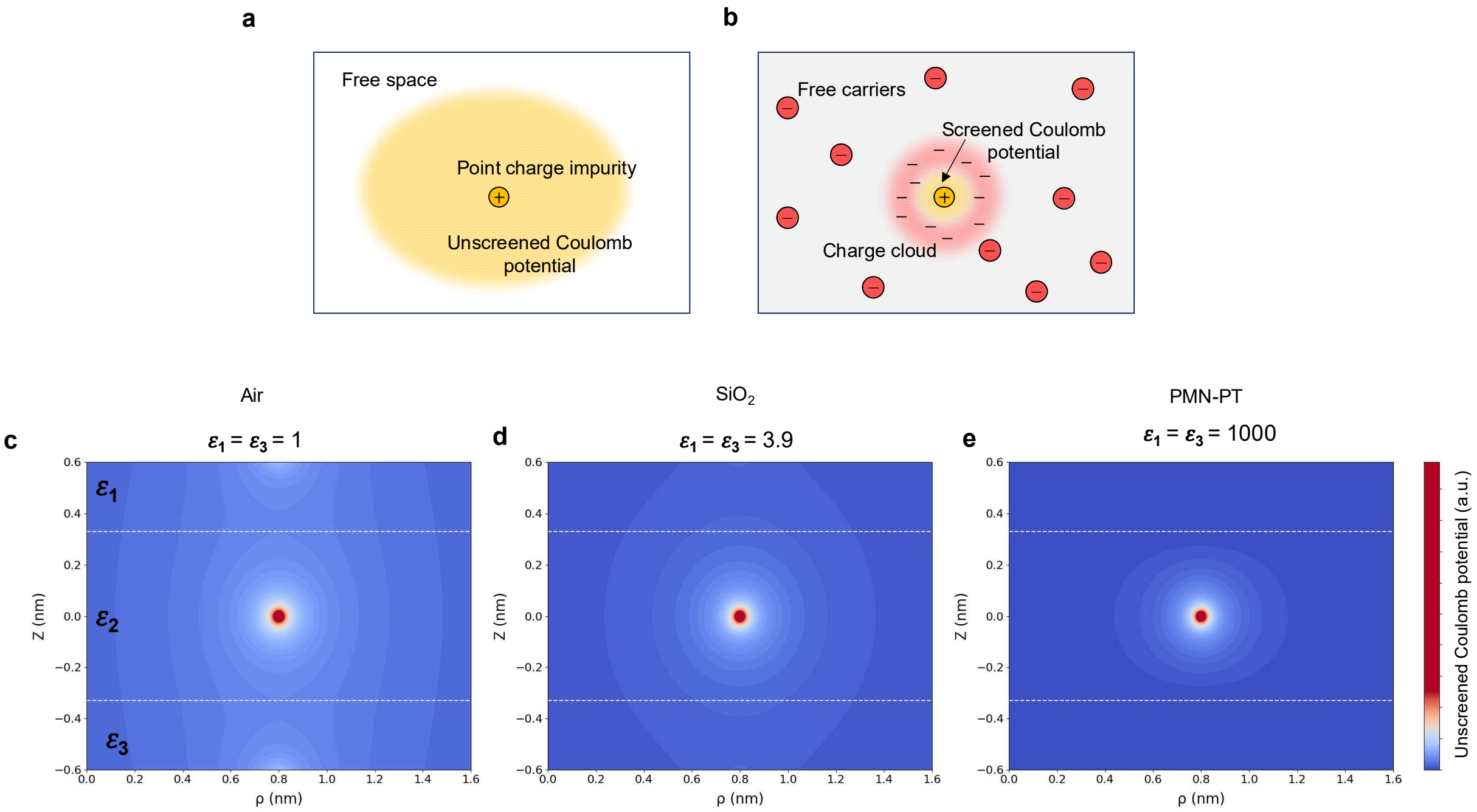

**Figure S1 | Effect of screening by dielectric environment**. (a-b) Coulomb potential for a charged impurity and its screening by the free charge cloud. (c-e) Simulated Coulomb potential due to a point charge impurity at the center of monolayer $MoS_2$ with different environments (air, $SiO_2$, and PMN-PT).

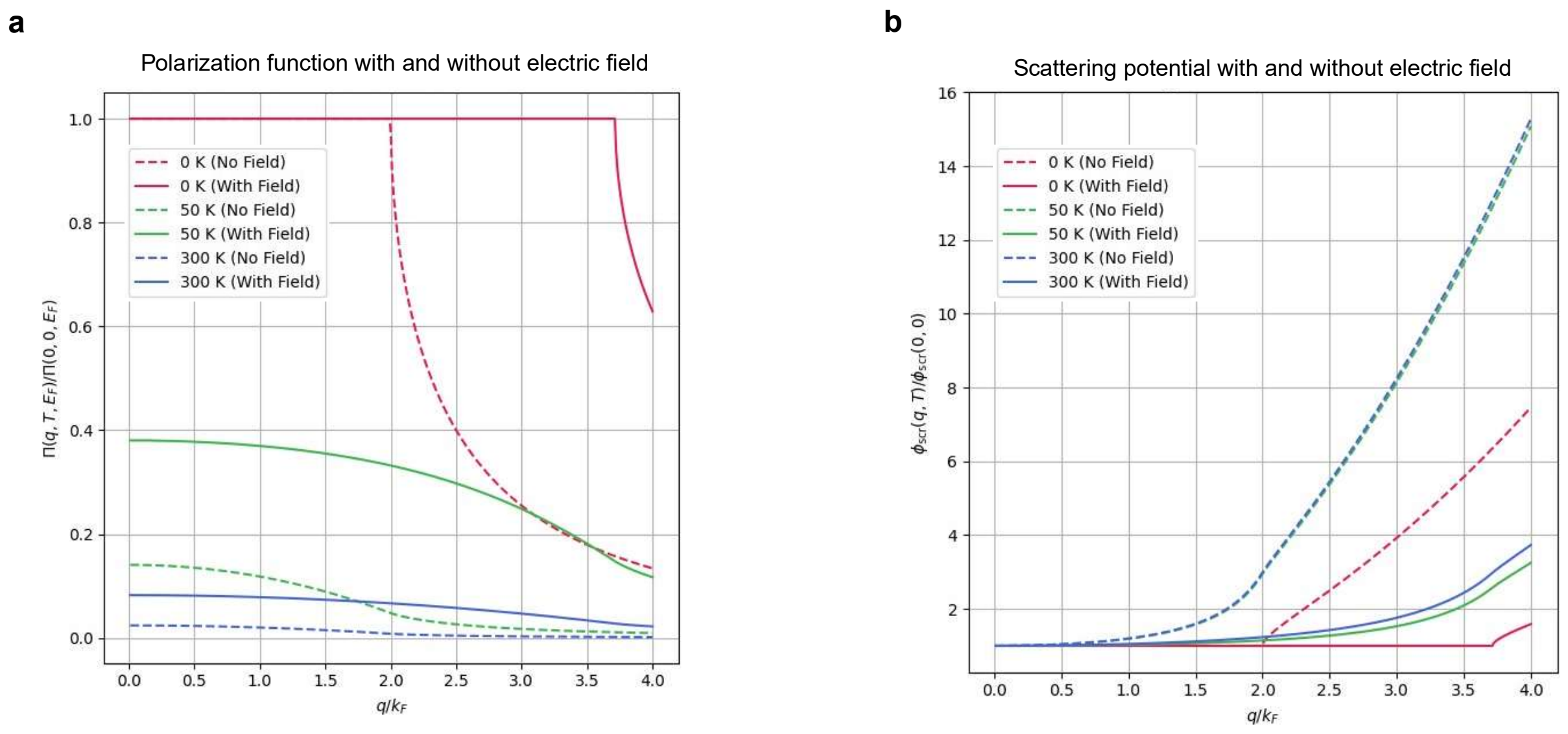

**Figure S2 | Reduced scattering *via* substrate polarization**. Normalized polarizability function (**a**) and scattering potential (**b**) as a function of momentum for HC-FET at 0 K (red), 50 K (green) and 300 K (blue), without (dashed) and with (solid) in-plane polarization derived from equation *(2.5)* and equation *(2.2)* in supplementary note section 1.2. The presence of an additional induced in-plane field helps to increase the polarizability of the carriers and hence the scattering potential is reduced even at room temperature. The enhancement of intrinsic carrier mobility is directly proportional to reduced scattering potential.

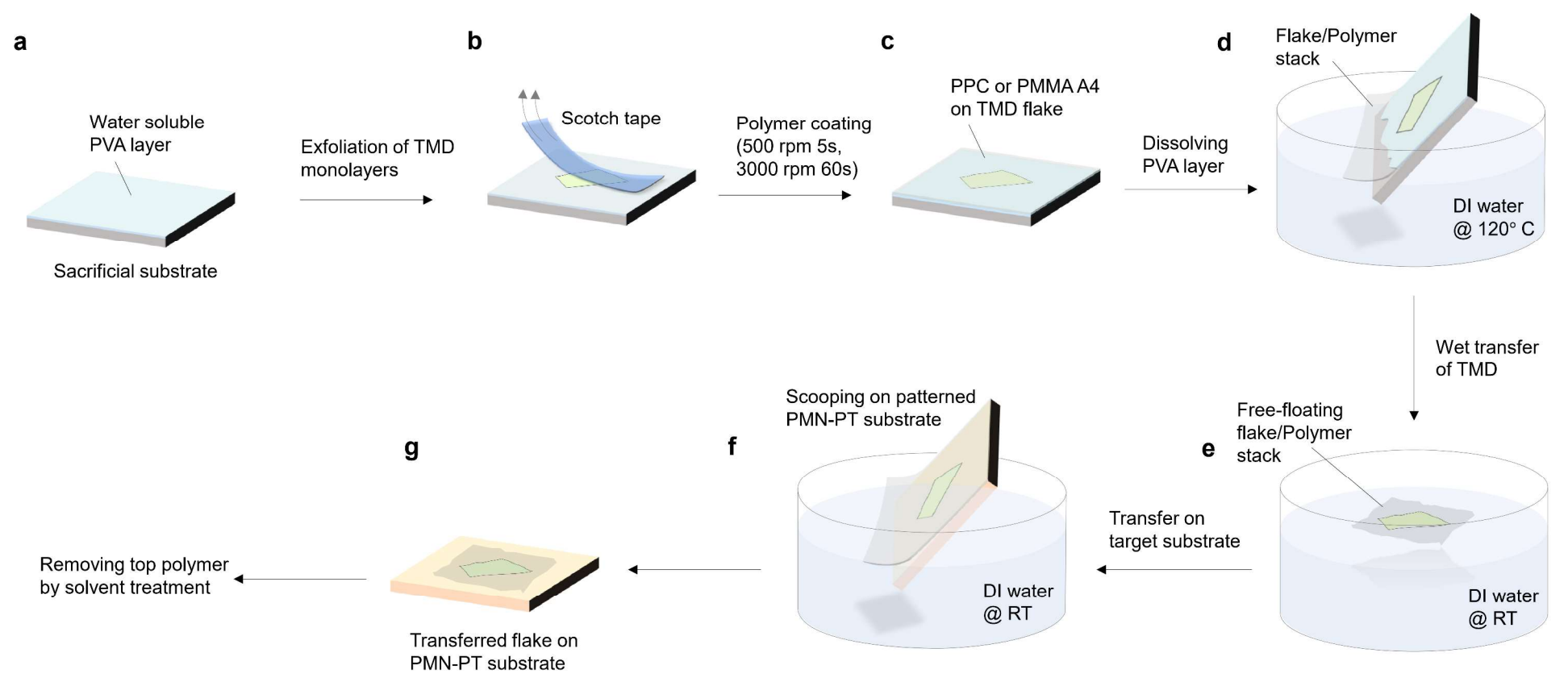

**Figure S3 | Device fabrication process**. (**a**) PVA layer was spin-coated on sacrificial $SiO_2$/Si where TMD flakes were mechanically exfoliated (**b**). Another polymer layer was coated on top for mechanical support (**c**). (**d-f**) The selected region of the flake/polymer layer was scooped on the target substrate by dissolving the PVA in hot water. (**g**) The substrate containing the target flake was dried up in ambient, followed by polymer removal in acetone, IPA and ethanol treatment.

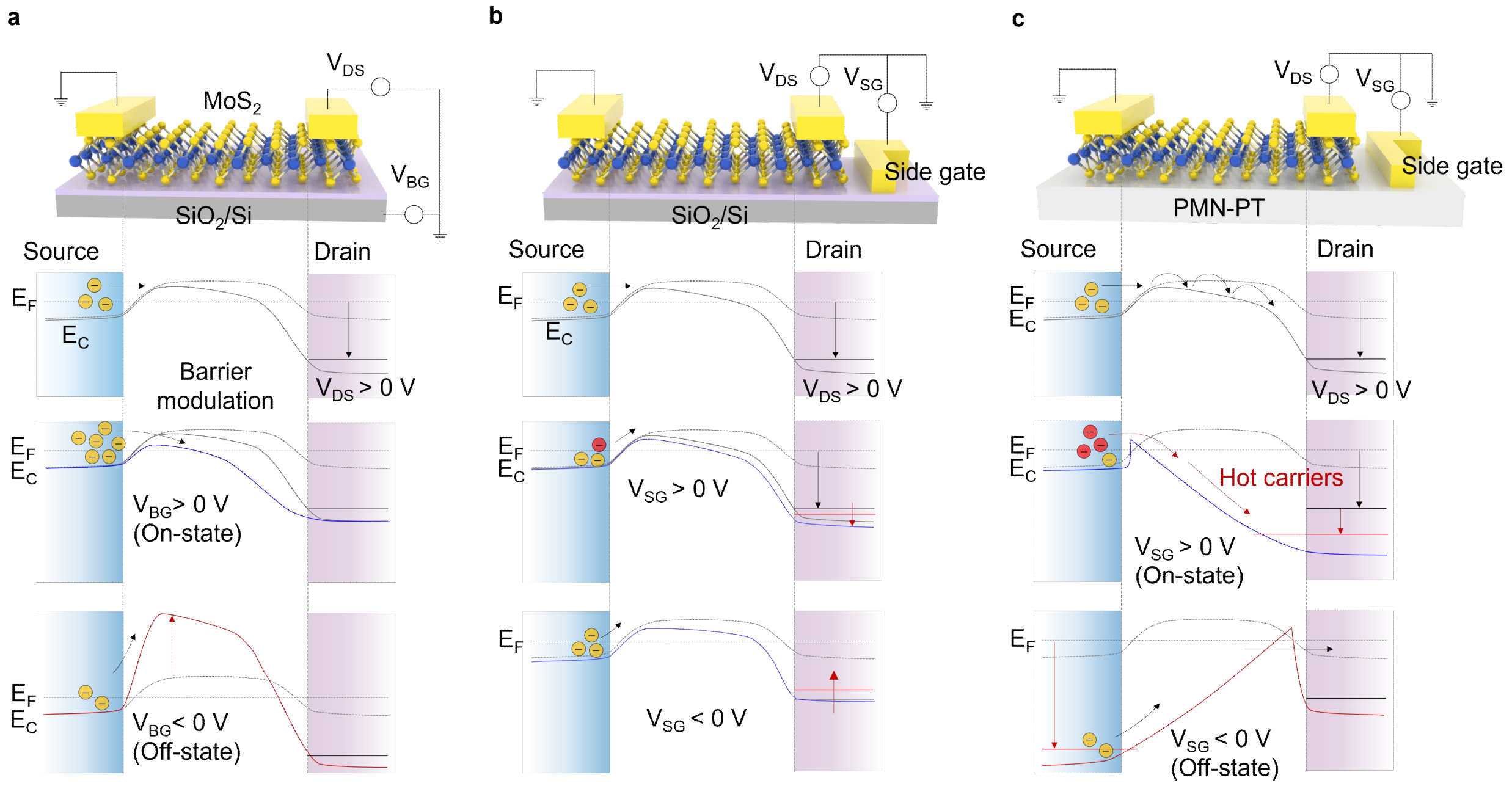

**Figure S4| Device schematics and band structures for various FET configurations**. Schematic device (top) and band structure (bottom) of n-type $MoS_2/SiO_2$ BG-FET (**a**), $MoS_2/SiO_2$ side-gate FET (SG-FET) (**b**), and HC-FET on PMN-PT substrate (**c**). The presence of strong in-plane polarization with side-gate leads to unconventional band modulation and hence hot-carrier generation.

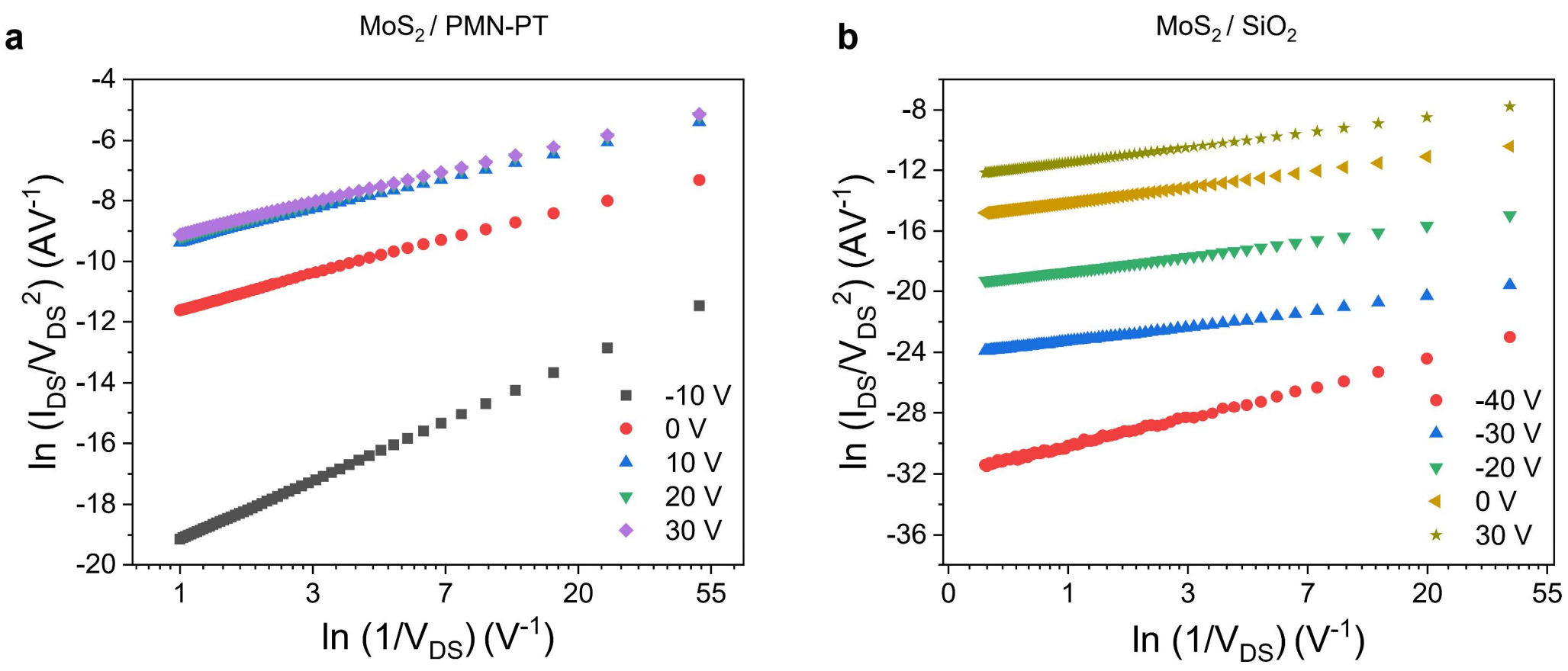

**Figure S5 | Ohmic contact with Bi/Au**. ln ($I_{DS}/V_{DS}^2$) plot as the function of inverse drain voltage. Plots indicate direct carrier tunneling in HC-FET (**a**) and BG-FET (**b**) under varying gate bias (mentioned in the respective plots).

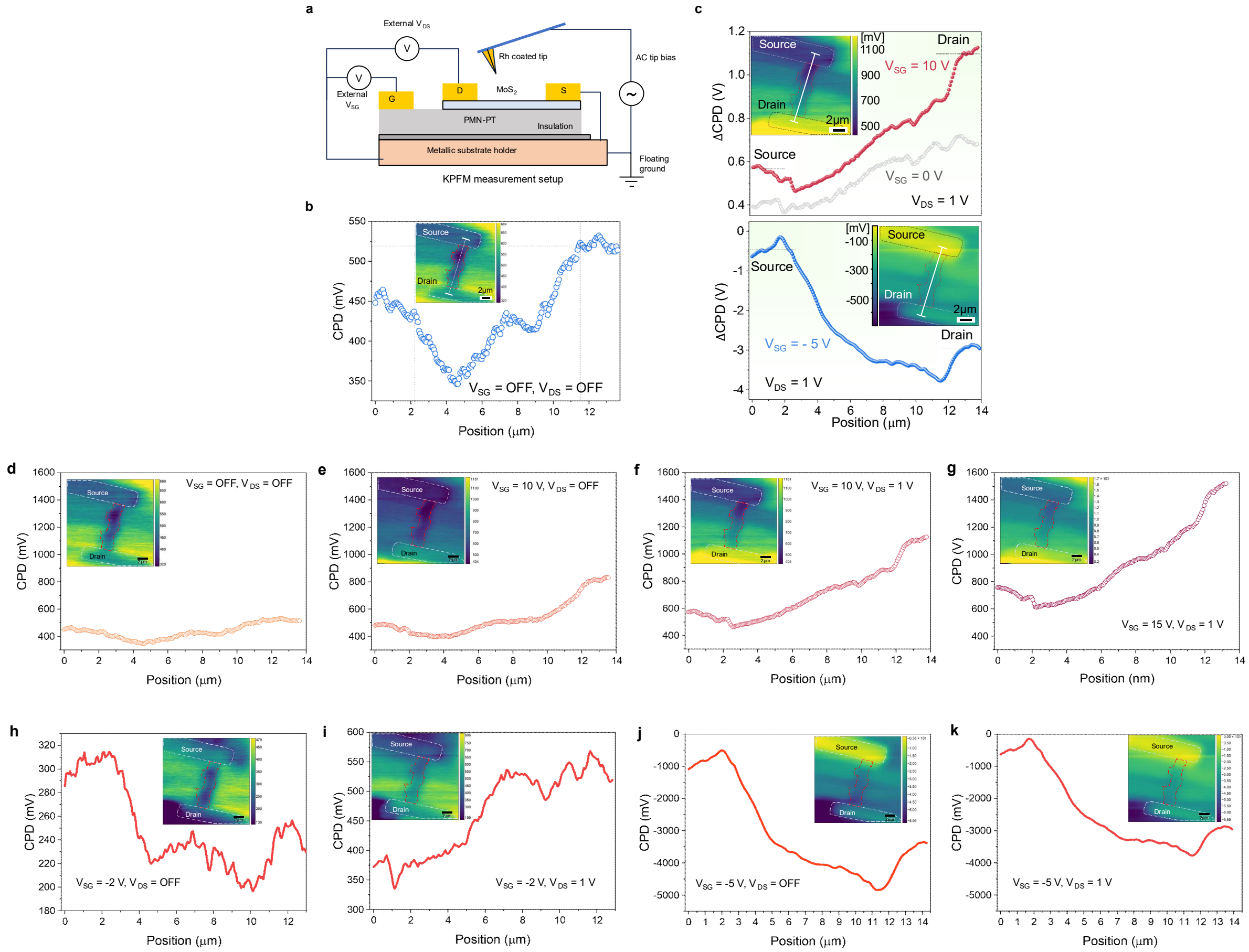

**Figure S6 | KPFM measurement**. (**a**) Schematic device setup for KPFM. (**b**) CPD plot without drain and gate bias. Inset shows potential maps for all the figures. (**c**) KPFM measurement of $MoS_2$ HC-FET. Measured CPDs at $V_{DS} = 1$ V are plotted for $V_{SG} = 0$ V (grey curve, top panel), $V_{SG} = 10$ V (red, top panel) and $V_{GS} = -5$ V (blue, bottom panel) referring to the drain voltage effect, FET on- and off-states, respectively. The CPD profiles are plotted along the white dashed lines. (d-g) The effect of gate (positive) and drain voltage in lateral band modulation. Corresponding $V_{SG}$ and $V_{DS}$ are mentioned in the respective plots. (h-k) Off-state CPD plots of HC-FET under negative gate bias.

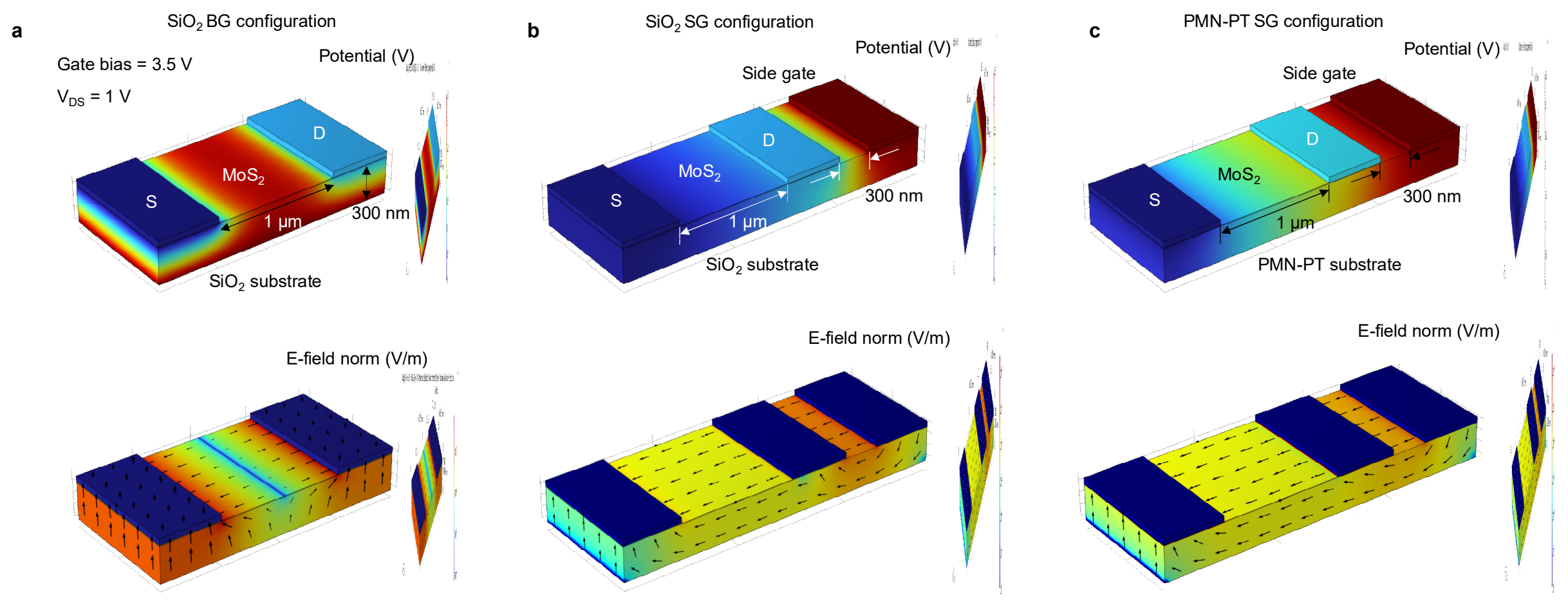

**Figure S7| Effect of in-plane polarization on carrier transport**. **a-c**, FEM simulation (at $V_{DS}$ = 1 V and gate bias of 3.5 V): Potential (top) and field (bottom) map for (**a**) typical $MoS_2$/ $SiO_2$ BG-FET,(**b**) $MoS_2$/$SiO_2$ SG-FET, and (**c**) $MoS_2$ HC-FET with side-gate electrode on PMN-PT.

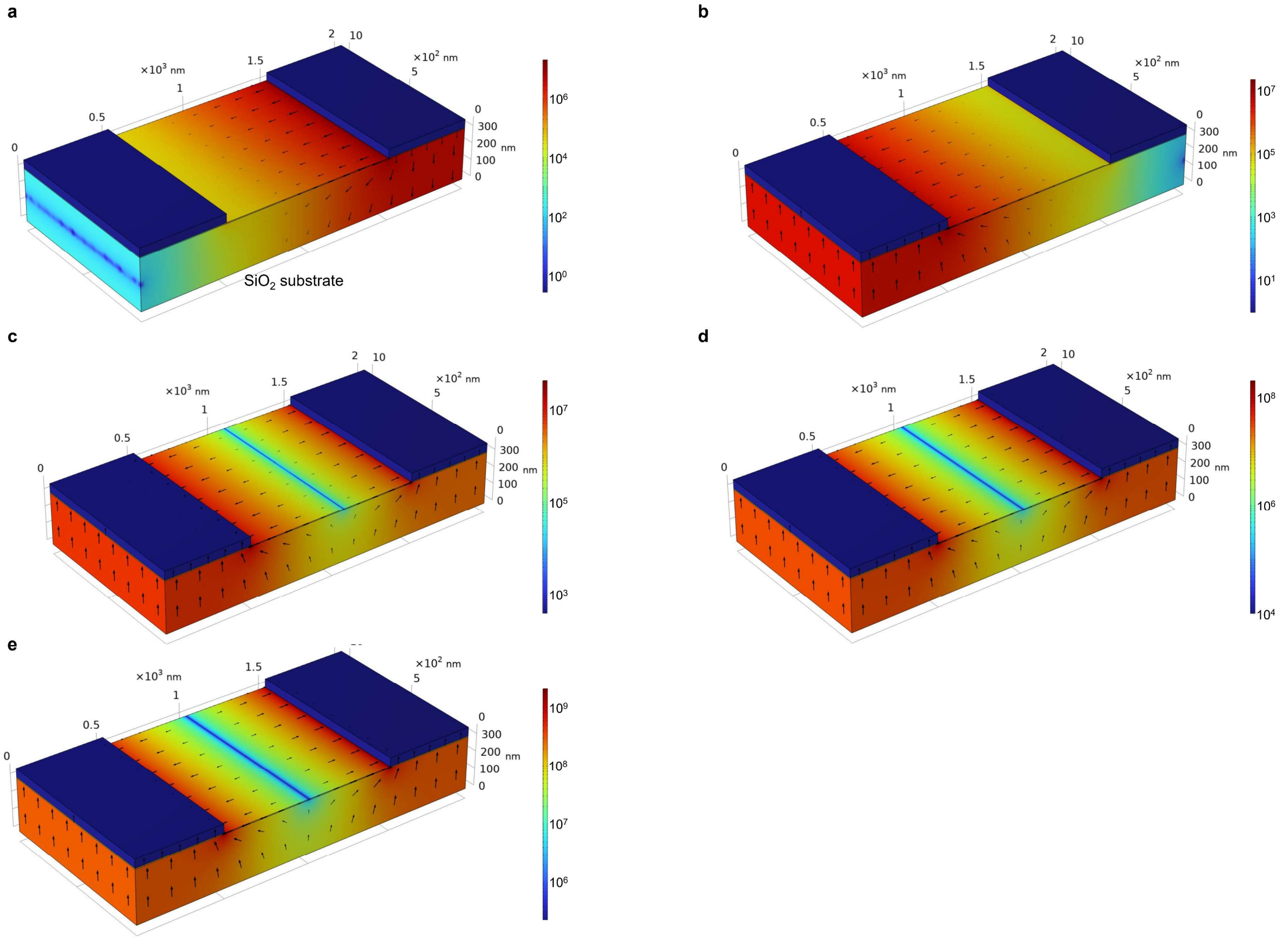

**Figure S8 | Parametric back-gate sweep for $MoS_2/SiO_2$ BG-FET**. (**a-e**) Electric field distribution under fixed $V_{DS}$ = 1 V, and $V_{BG}$ of 0 V, 1 V, 2 V, 10 V, and 100 V. The direction of the electric field is represented by arrows.

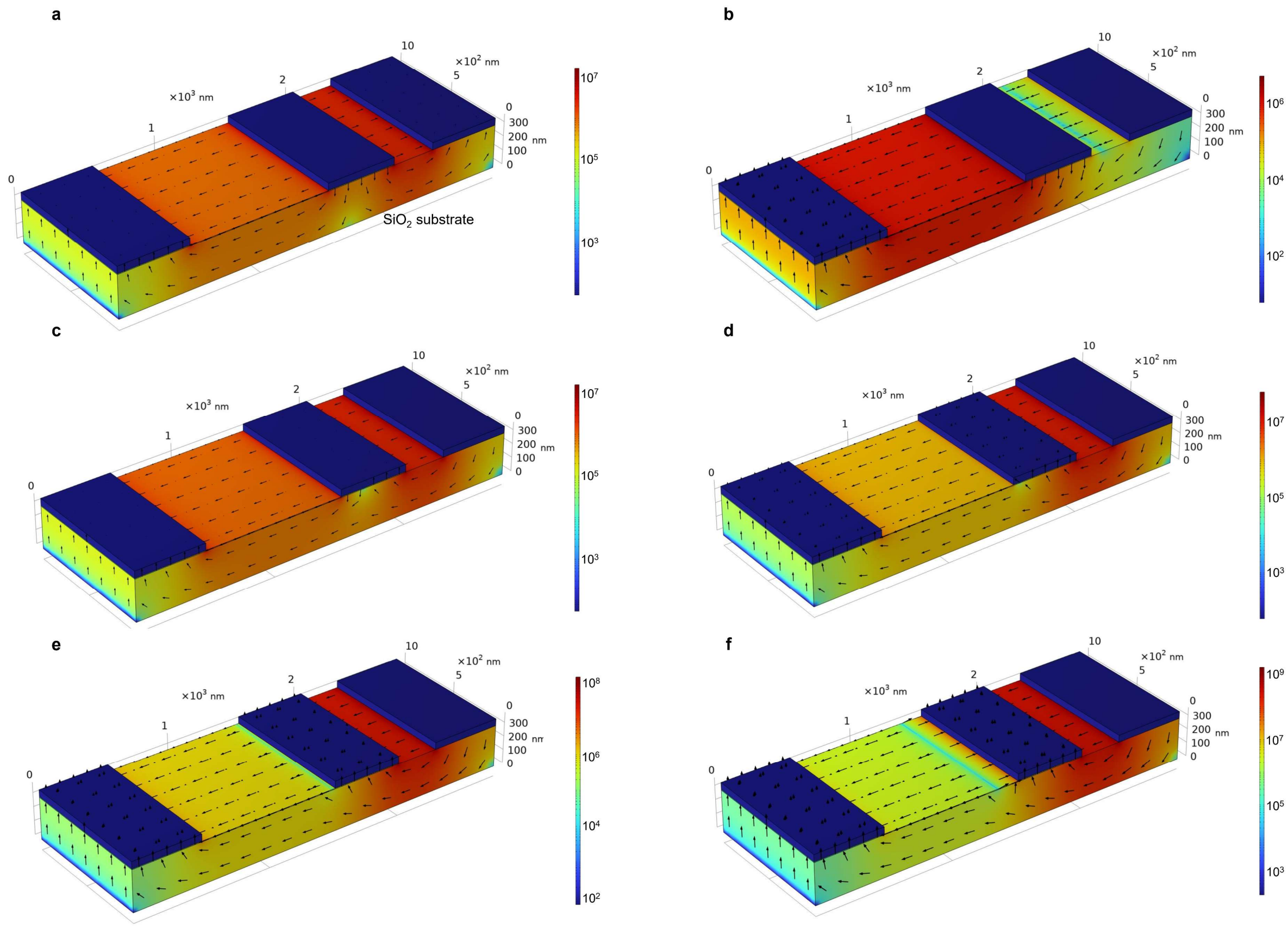

**Figure S9 | Parametric sweep for $MoS_2/SiO_2$ SG-FET**. A side-gate is positioned 300 nm from the channel. All the material parameters are retained identical to the BG-FET. (**a-f**) Norm of electric field distribution under fixed $V_{DS}$ = 1 V, and $V_{SG}$ of 0 V, 1 V, 2 V, 5 V, 10 V, and 100 V. The direction of the electric field is represented by arrows.

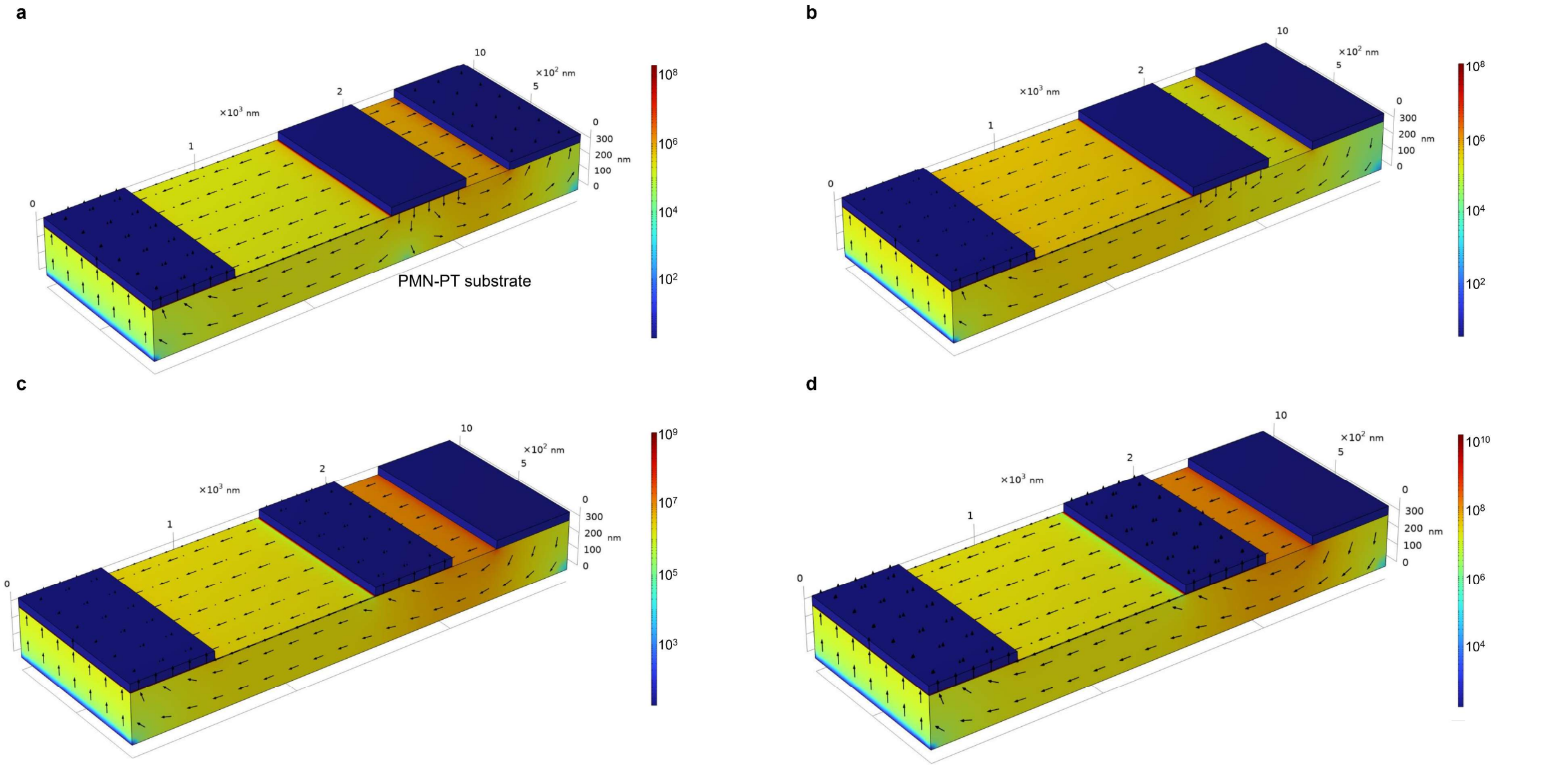

**Figure S10 | Parametric sweep for $MoS_2$/PMN-PT HC-FET**. A side-gate electrode is positioned 300 nm from the channel and the device dimensions are kept the same as FETs on $SiO_2$ for direct comparison. (**a-d**) Norm of electric field distribution under fixed $V_{DS}$ = 1 V, and $V_{SG}$ of 0 V, 1 V, 10 V, and 100 V. The direction of the electric field is represented by arrows.

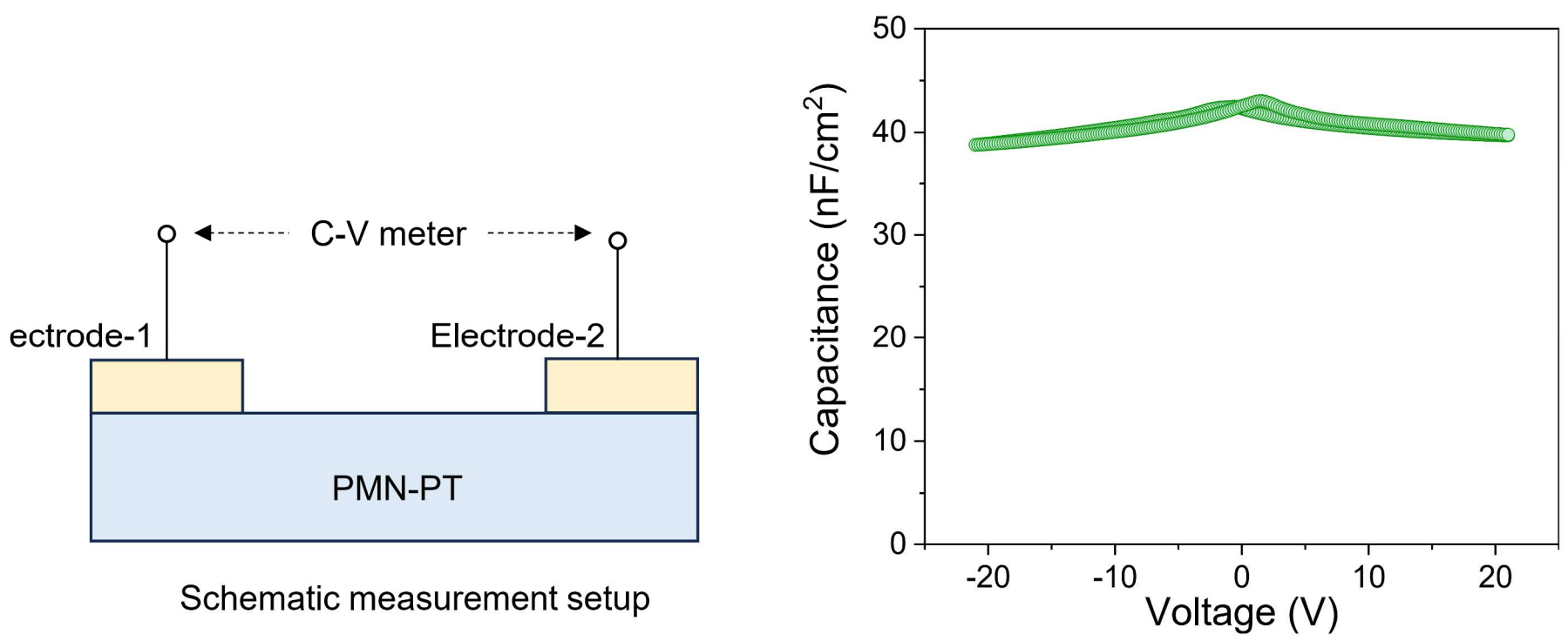

**Figure S11 | Capacitance measurement**. Schematic of the measurement setup (Left) and capacitance per unit area as the function of drive voltage (Right) is plotted for PMN-PT in-plane capacitor.

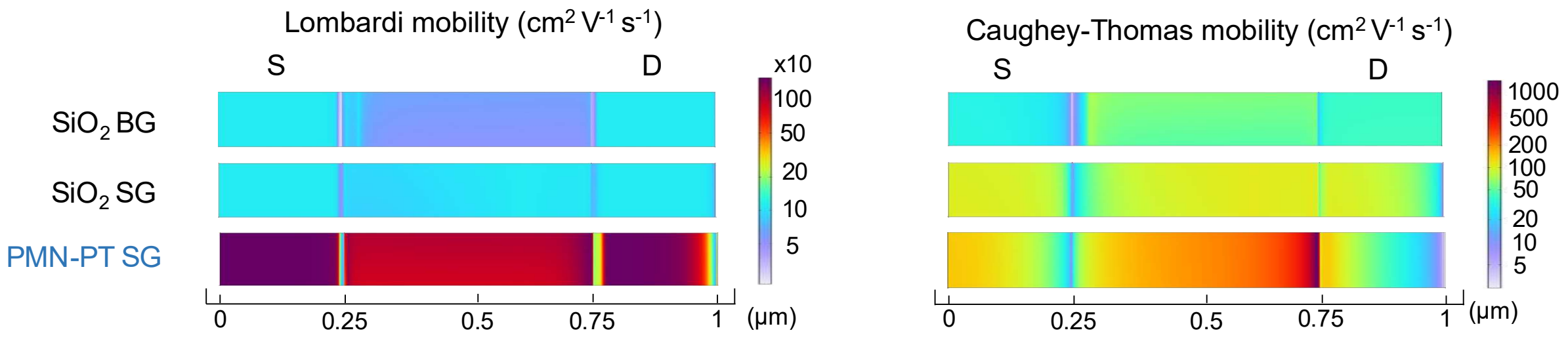

**Figure S12| Effect of in-plane polarization on carrier mobility**. Simulated Lombardi surface mobility after including acoustic phonon and surface roughness scatterings (top) and Caughey-Thomas mobility considering optical phonon scattering and high field velocity saturation (bottom) for $MoS_2/SiO_2$ BG-FET, $MoS_2/SiO_2$ SG-FET, and $MoS_2$/PMN-PT HC-FET.

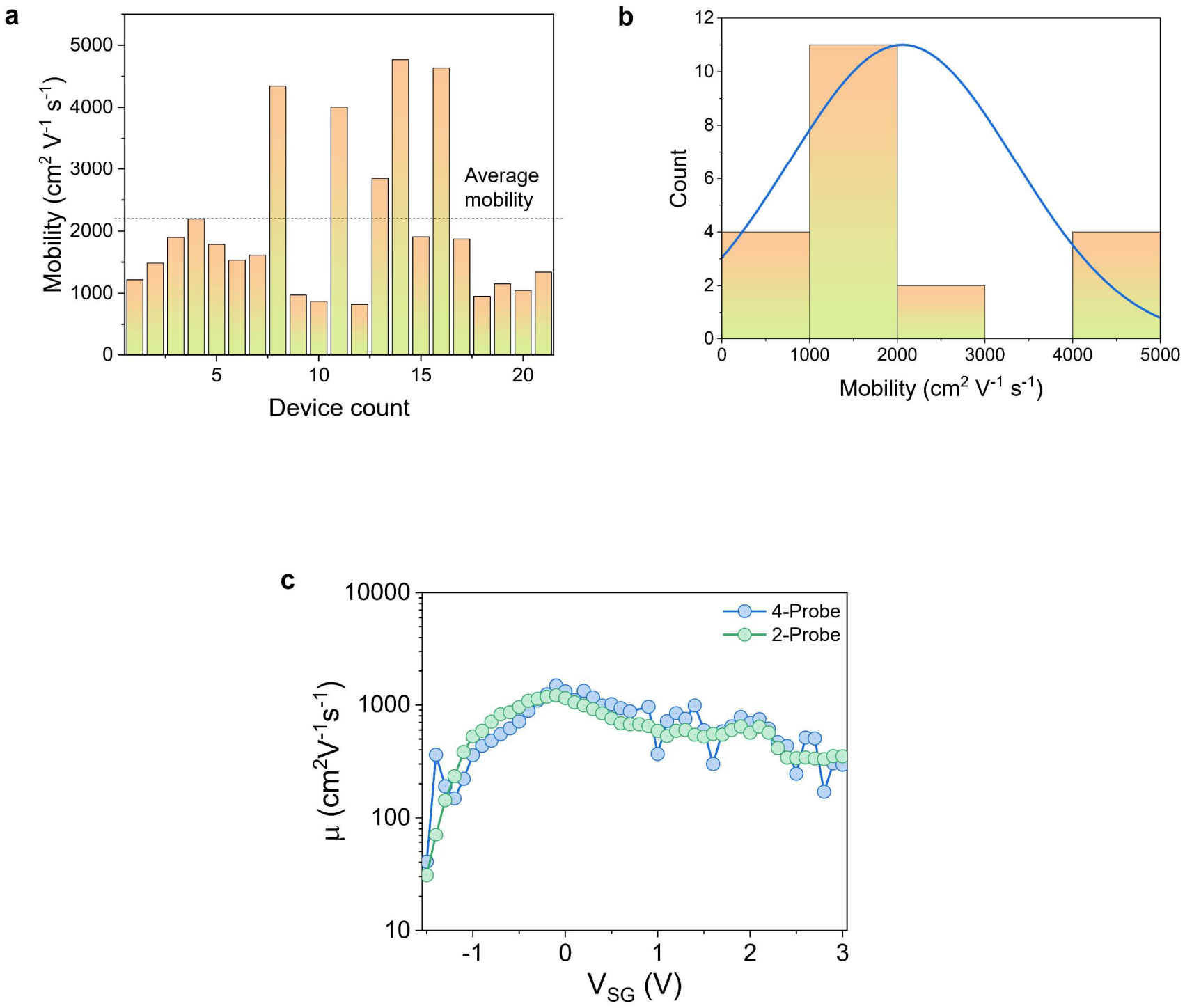

**Figure S13 | Performance reproducibility with various HC-FETs**. (**a-b**) Statistical distribution of mobility for twenty-one different $MoS_2$ HC-FETs is provided, yielding an average mobility of ~2200±10 $cm^2V^{-1}s^{-1}$. (**c**) Mobility obtained from simultaneously measuring two- and four-probe $I_{DS}$-$V_{SG}$ are plotted as the function of gate bias.

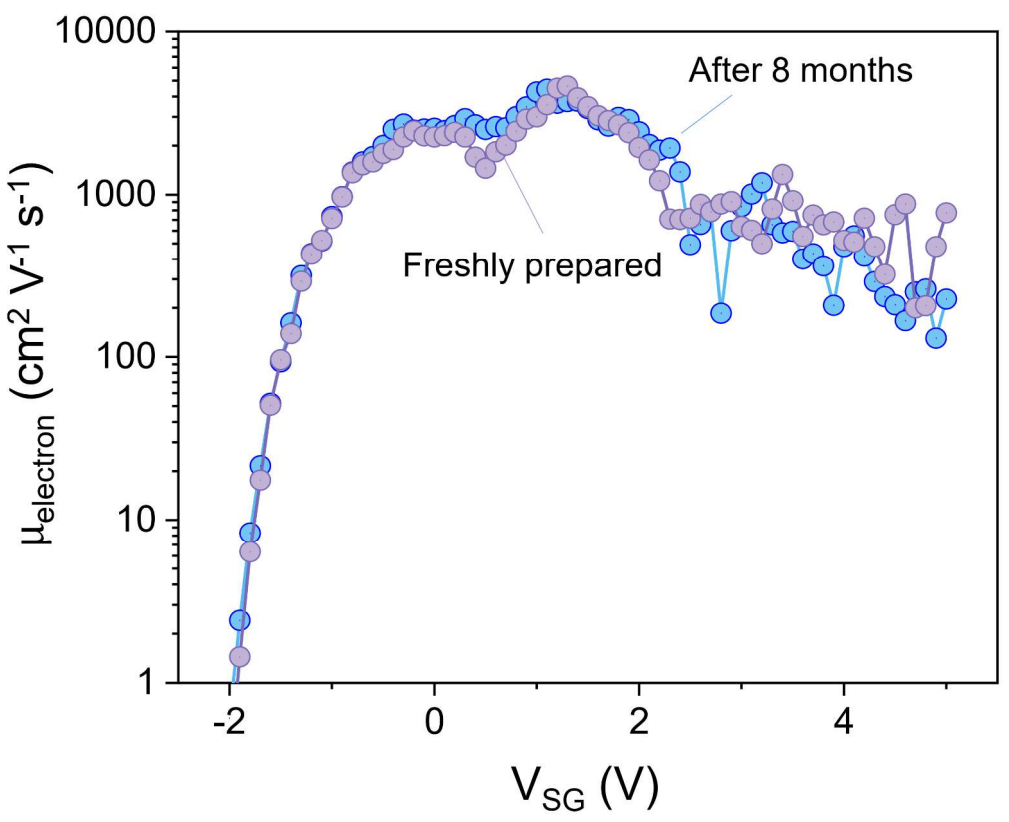

**Figure S14 | Aging effect investigation**. The performance stability was tested over an extended period (after 8 months) indicating good agreement with the performance of freshly prepared samples.

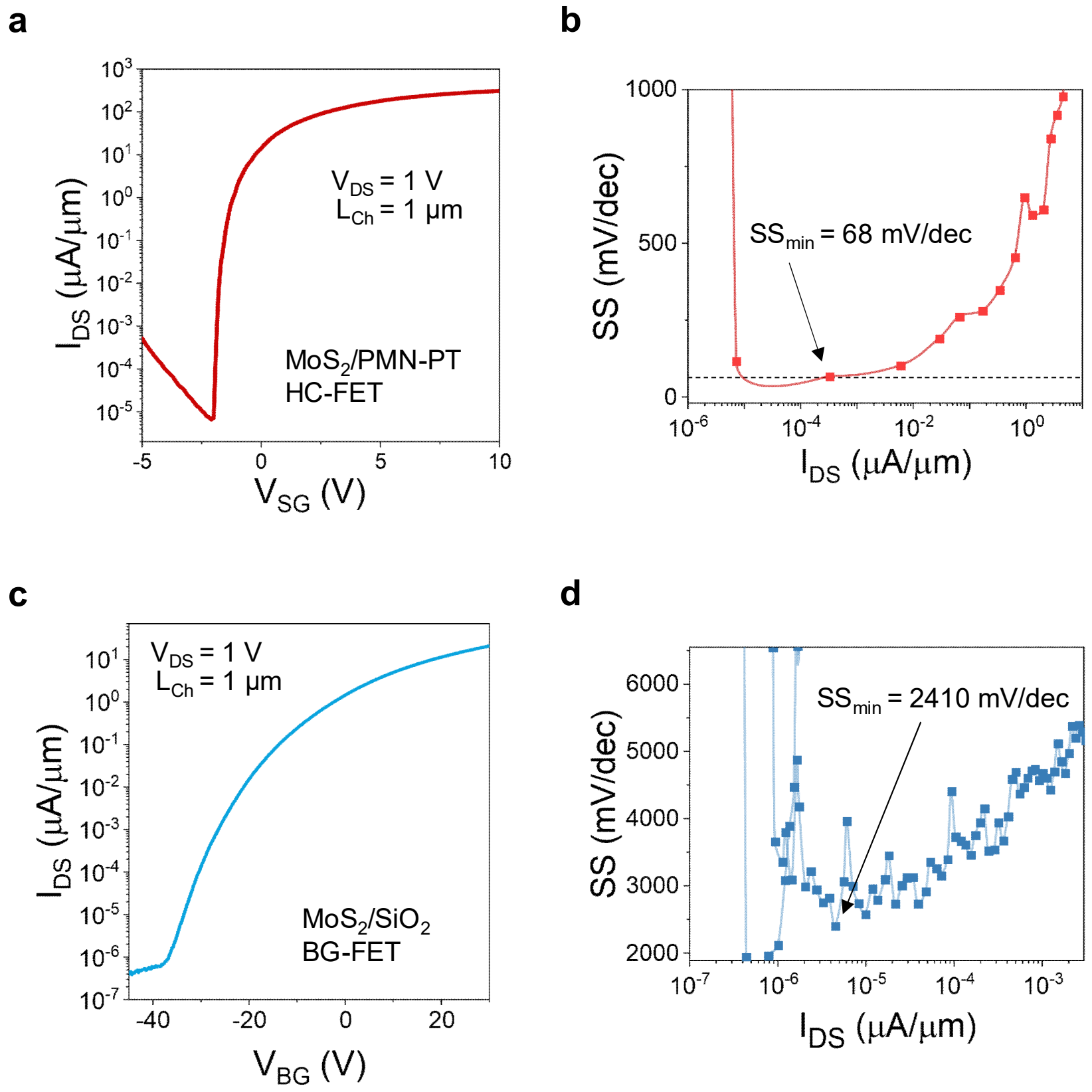

**Figure S15 | Subthreshold swing (SS) performance**. (**a**) Transfer characteristics of n-type HC-FET. (**b**) The SS as a function of drain current with a minimum (68 mV/dec) value reaching nearly the thermal limit. (**c**) Transfer characteristics of standard back gated n-FET of identical dimensions. (**d**) The SS *vs* drain current shows minimum SS of 2410 mV/dec.

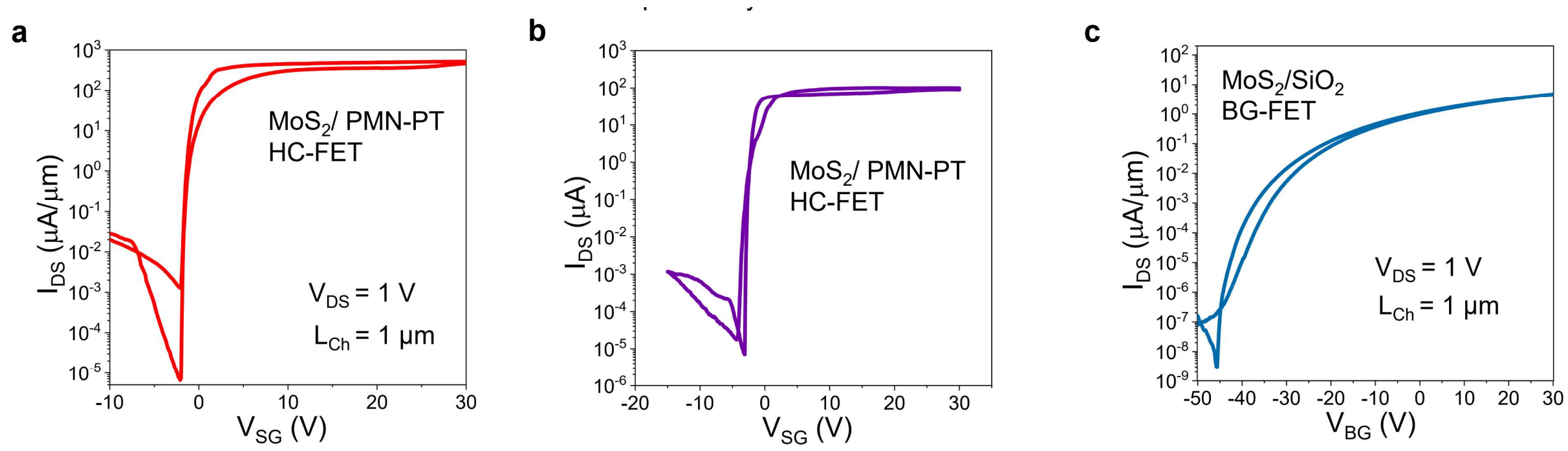

**Figure S16 | Hysteresis in $MoS_2$ FET**. (**a**) Despite the presence of ferroelectric substrate, dual sweep transfer curve of a HC-FET shows negligible hysteresis. (**b**) Similar nominal hysteresis behavior was reproduced in another HC-FET. (**c**) Dual sweep transfer characteristics in BG-FET with $SiO_2$ show much larger hysteresis than HC-FET. This confirms that the observed hysteresis is associated with the trap-states which are efficiently screened in HC-FETs.

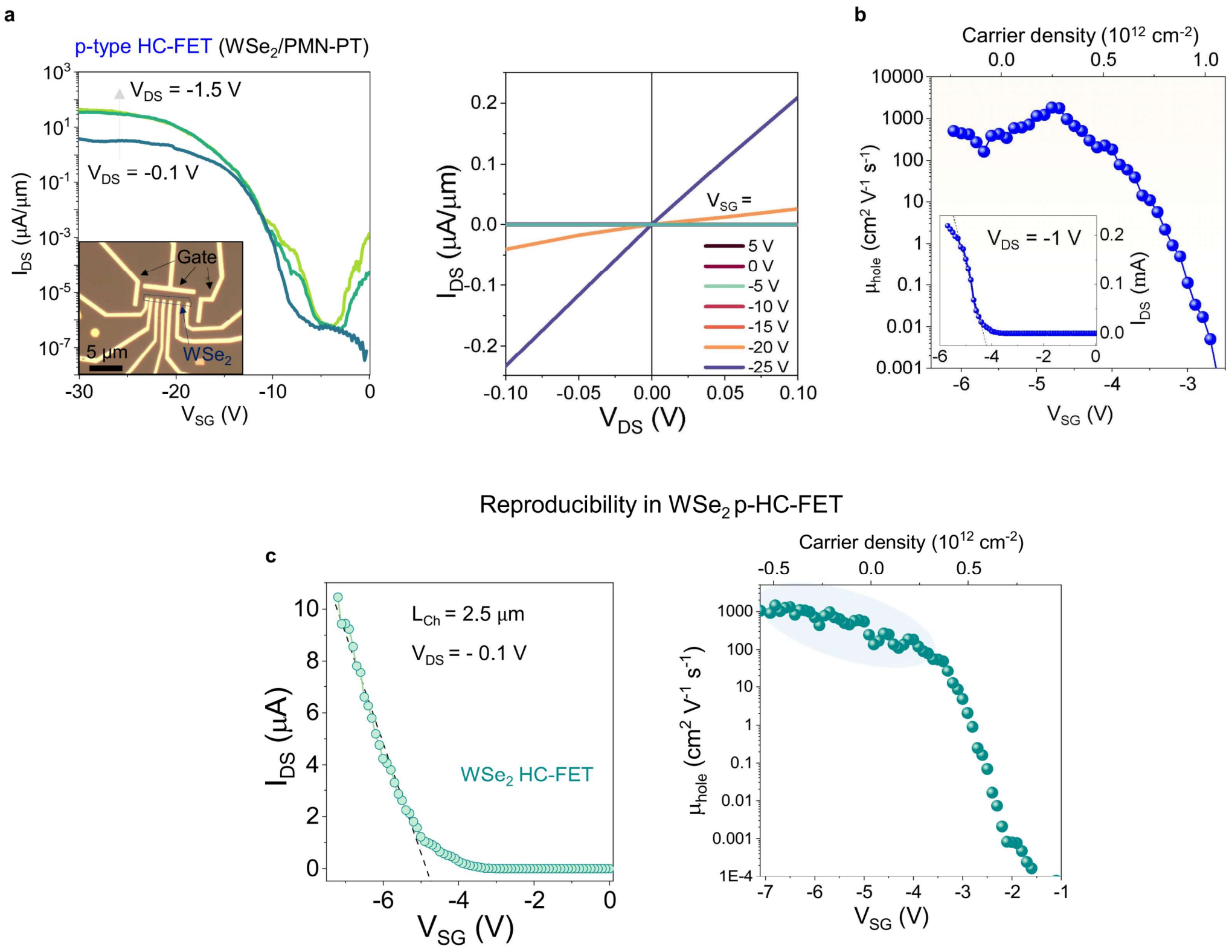

**Figure S17 | p-type HC-FET**. (**a**) Transfer curve of monolayer p-type $WSe_2$ HC-FET. (**b**) Measured mobility as a function of gate bias. (**c**) Performance reproducibility in another $WSe_2$ HC-FET.

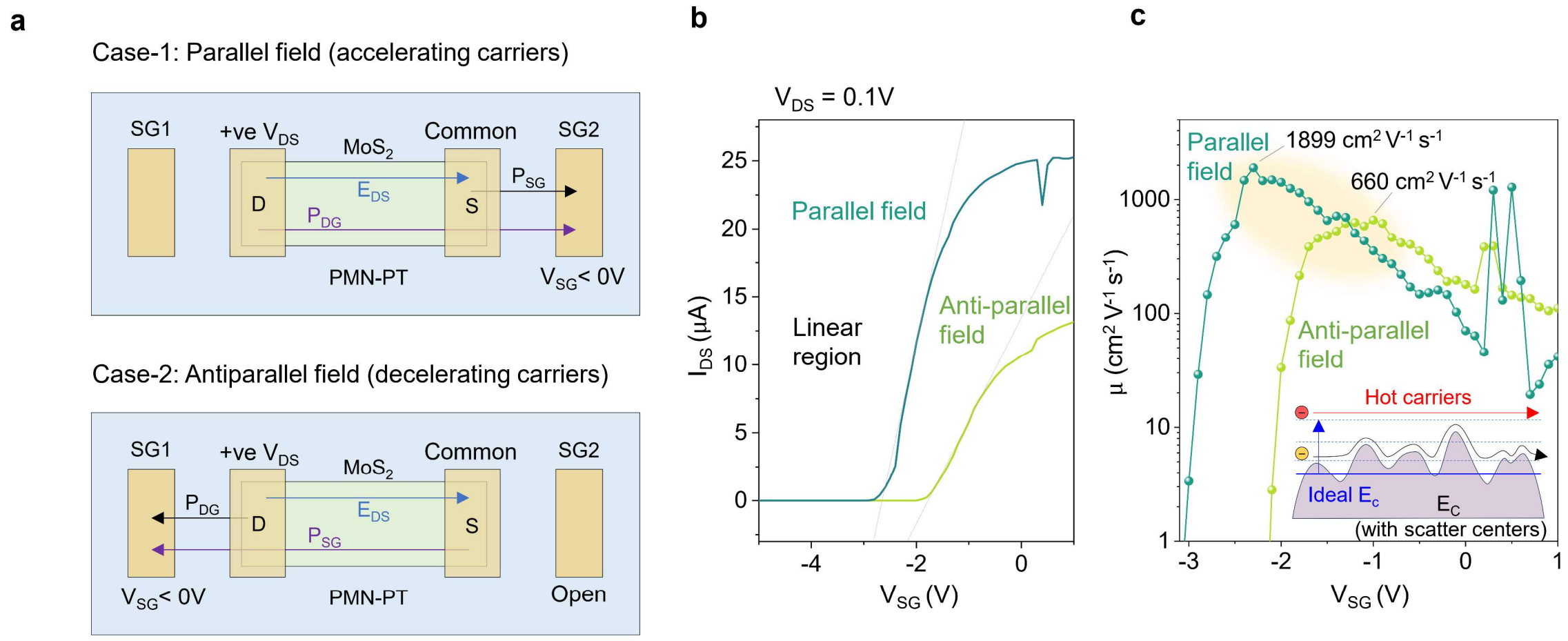

**Figure S18| Effect of in-plane polarization on carrier transport**. **(a)** Schematic illustration of HC-FET with dual side-gates (SG1 and SG2) which are positioned equidistant from the channel. The gates can be configured for parallel (top) or anti-parallel (bottom) polarization. **(b)** Transfer characteristics of the same device under parallel and anti-parallel gate configuration. **(c)** Corresponding FET mobility changes with gate bias. The inset shows hot-carrier mechanism under an accelerating field.

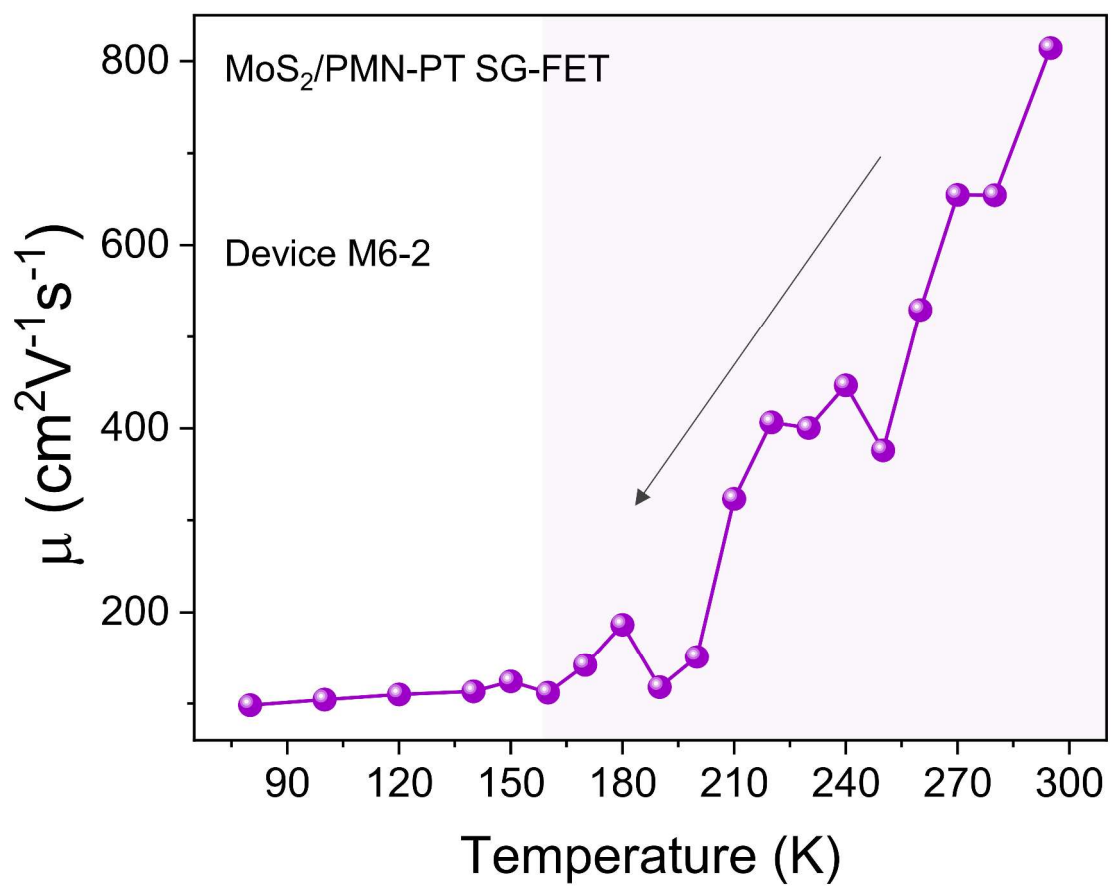

**Figure S19 | Temperature dependence of mobility**. Reproducible mobility changes as the function of temperature for one of the HC-FETs (Device M6-2).

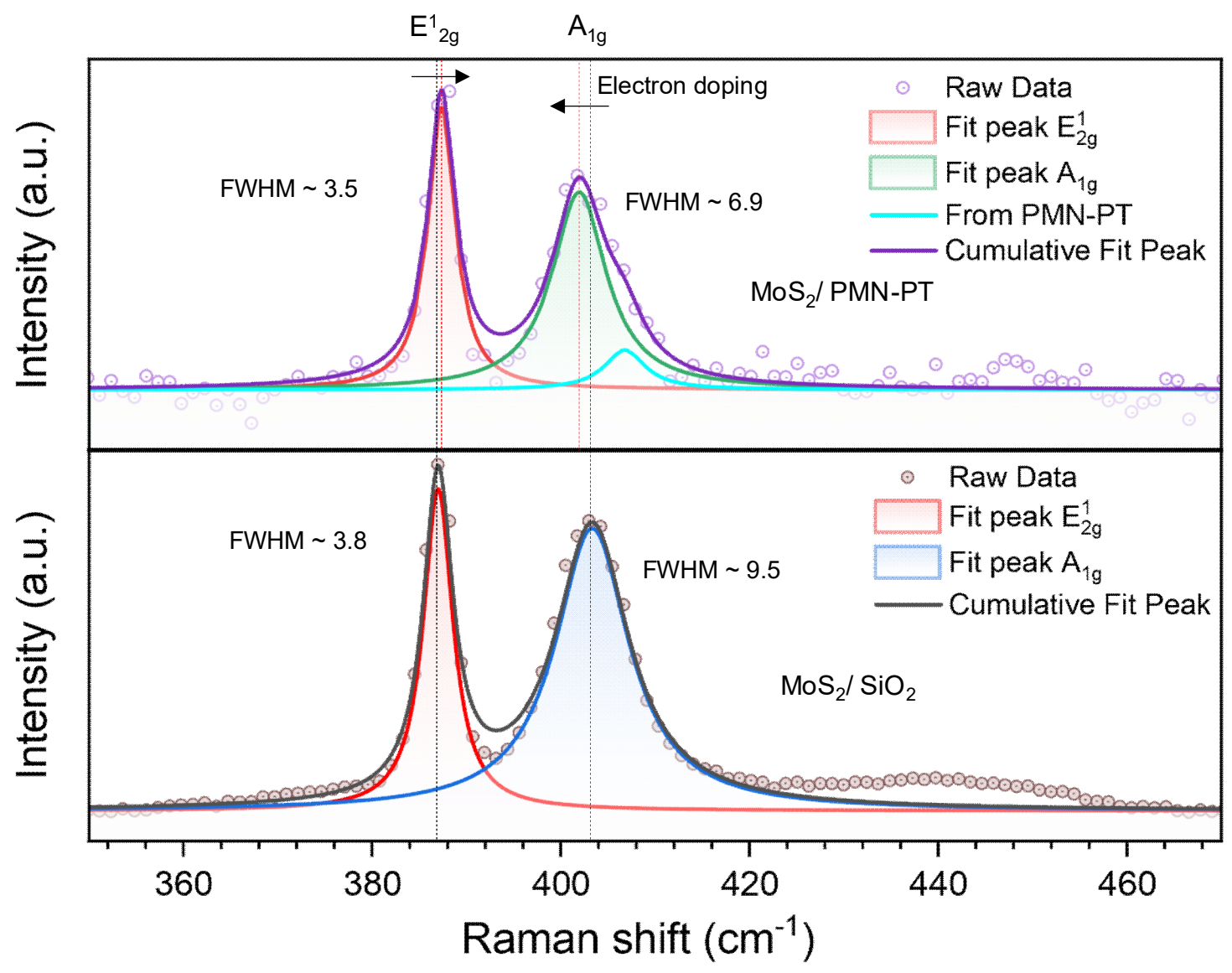

**Figure S20 | Raman Spectroscopy**. Deconvoluted Raman spectra of $MoS_2$ on PMN-PT substrate (top) and on $SiO_2$ substrate (bottom).

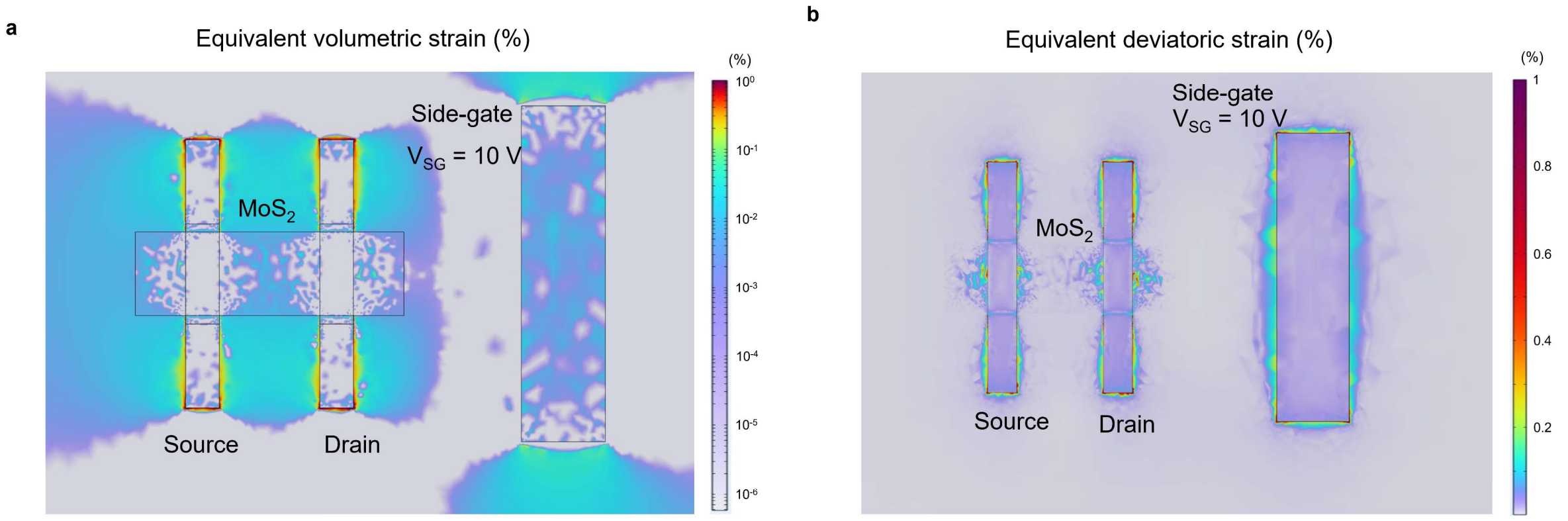

**Figure S21 | Strain mapping in HC-FET due to inverse-piezoelectric effect of PMN-PT substrate**. (**a**) Equivalent volumetric strain and (**b**) deviatoric strain mapping at $V_{DS}$ = 1 V and $V_{SG}$ = 10 V. Both strains remain far below 1%, indicating no significant enhancement of mobility induced by strain in HC-FETs.

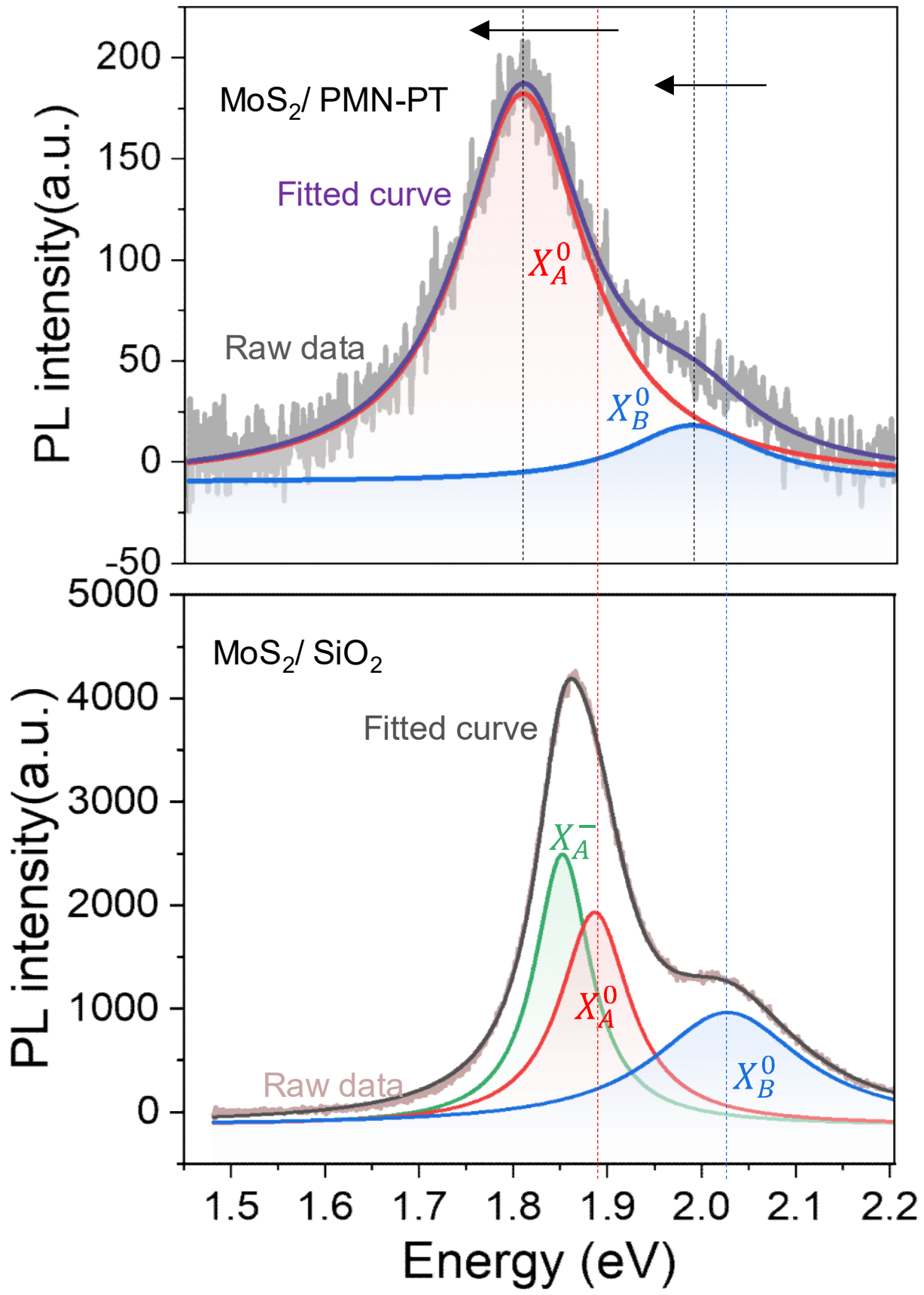

**Figure S22 | PL spectroscopy of $MoS_2$ on PMN-PT**. Deconvoluted PL spectra on PMN-PT (top) and on $SiO_2$ (bottom). The $MoS_2$/PMN-PT sample exhibits no trion peak, while the trion peak is dominant on $SiO_2$.

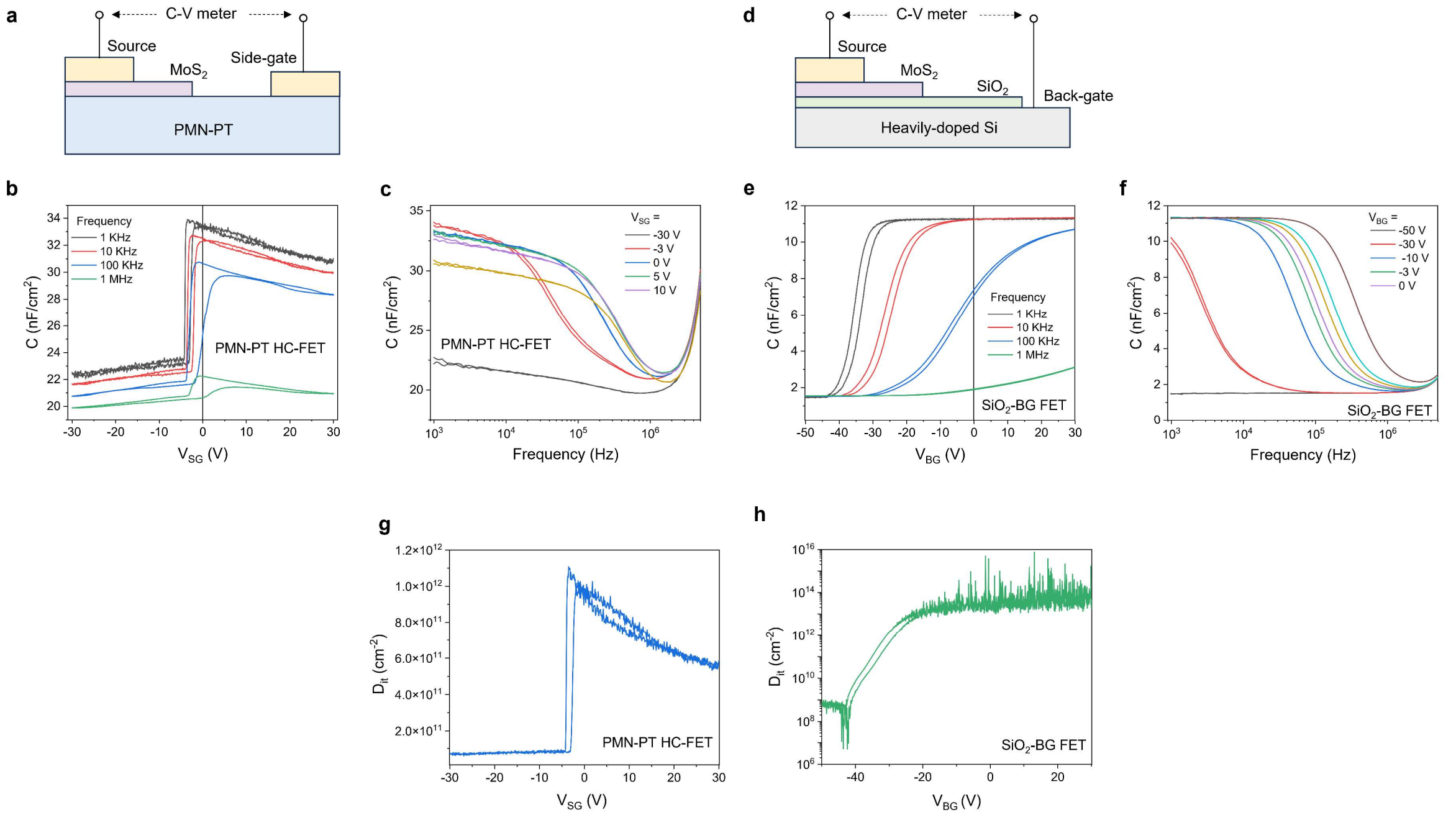

**Figure S23 | Trap density extraction from capacitance measurement**. Schematic *C-V* measurement setup and gate bias and frequency-dependent metal-insulator-semiconductor capacitance measured for HC-FET (**a**-**c**) and standard BG-FET (**d**-**f**). The trap density is estimated by standard high-low frequency method for HC-FET (g) and BG-FET (**h**). In HC-FET, effects of the traps are efficiently reduced owing to the strong dielectric screening.

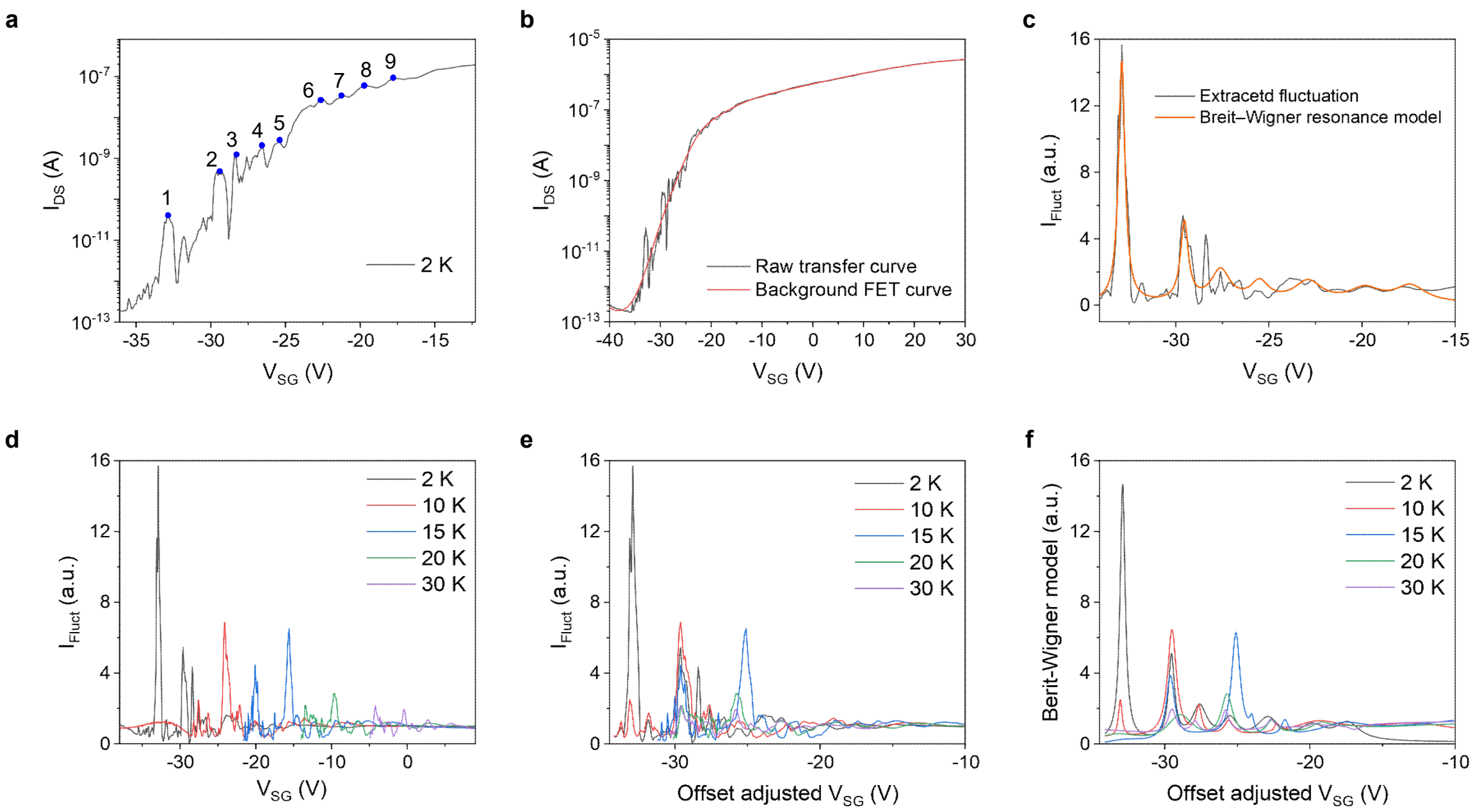

**Figure S24| Resonant tunneling through localized states**. (**a**) Marked fluctuation peaks on transfer curve (2K), reproducibly surviving across temperatures. (**b**) Subtracting classical FET background. (**c**) Fitting the extracted fluctuation component with Breit-Wigner resonance current model. (**d**) Extracted fluctuation components across different temperatures. (**e**) Off-set adjusted fluctuation components free of any classical transport contributions. (**f**) Simulated fluctuation components from Breit-Wigner model resembling the experimental fluctuation behavior.

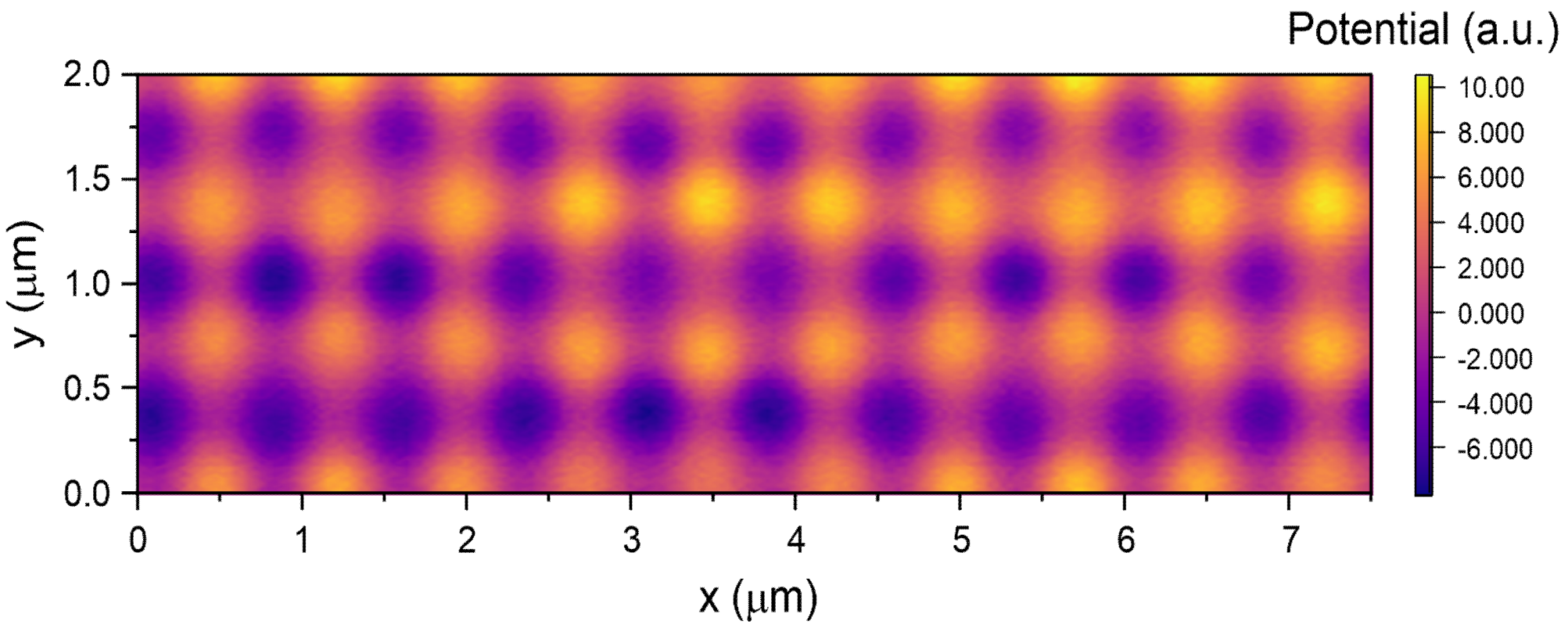

**Figure S25| Simulated potential landscape at 2K**. Quasi-periodic potential profile obtained by solving coupled 2D Poisson-Fermi-Thomas equations incorporating e-e interactions, spatially varying screening and boundary conditions relevant to device geometry.

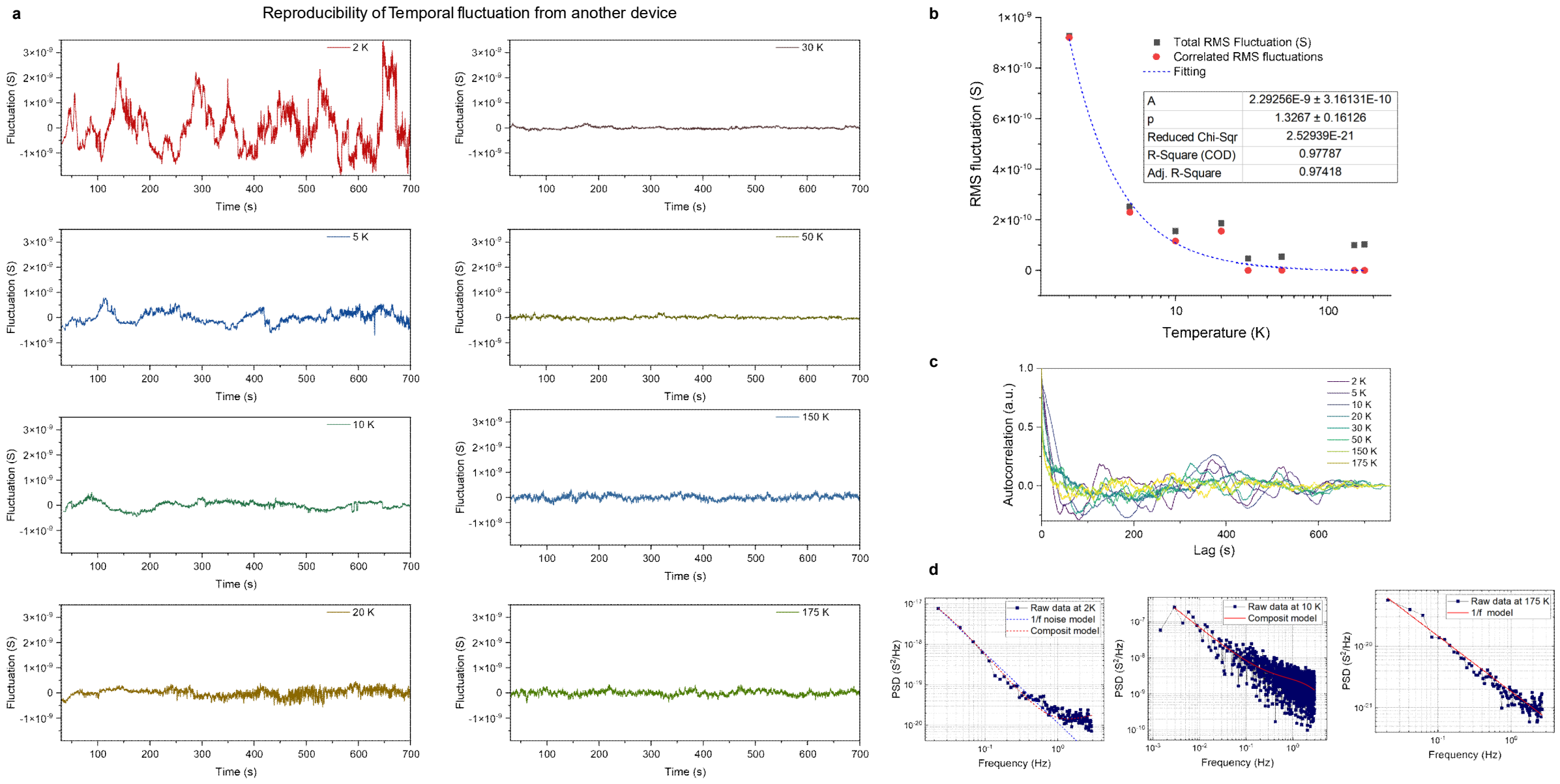

**Figure S26| Temporal fluctuations**. (**a**) Reproducible temporal fluctuations across a wide temperature window. (**d**) Gradual decay in RMS fluctuation *vs* temperature confirms thermal smearing effect. (**c**) Autocorrelation functions estimated at different temperatures show gradual loss of correlation memory with temperature. (**d**) PSD at three representative temperatures (2 K, 10 K, and 175 K). At higher temperature along with the fitting curves. The PSD can be well explained by flicker noise at 175 K while at lower temperature correlation effects take over leading to distinct noise behavior.

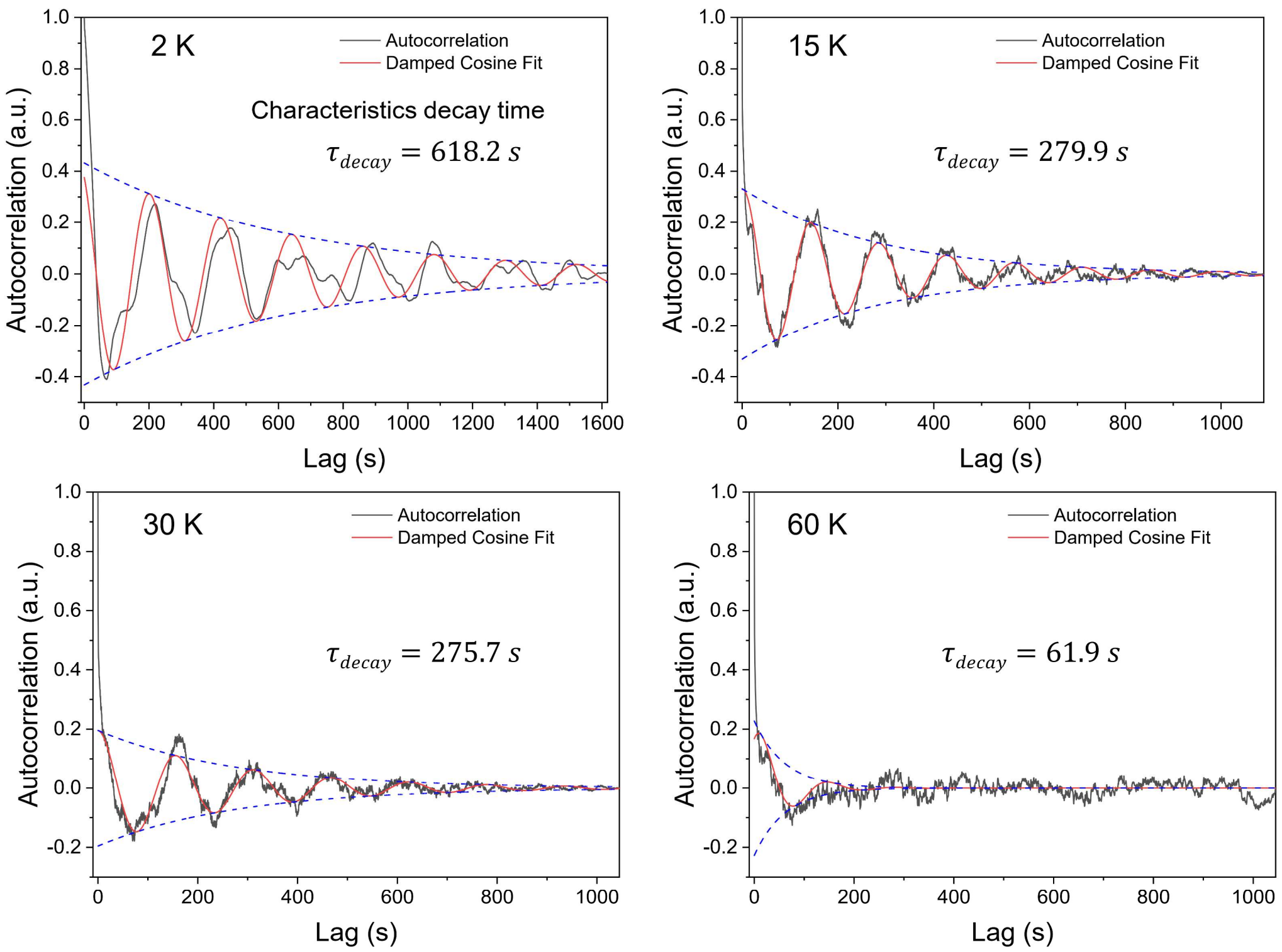

**Figure S27 | Characteristic decay time analysis**. Decay time at different temperatures is estimated from Fig. 3g autocorrelation datasets by logarithmic envelop function fitting. With increasing temperature, autocorrelation decays faster, attributing to thermal delocalization.

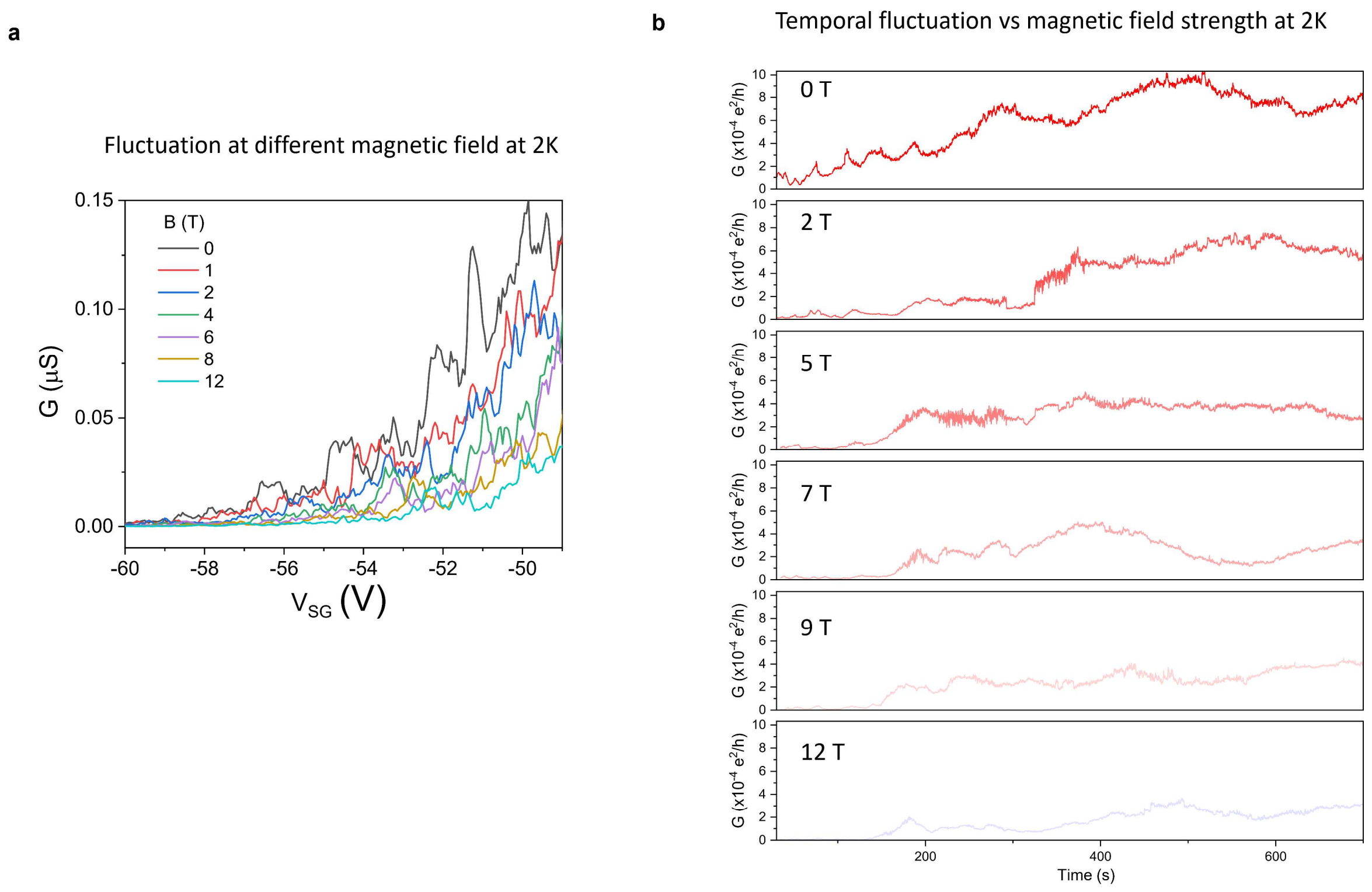

**Figure S28 | Effect of magnetic field on the correlated phase**. (**a**) With increasing B-field, the fluctuation amplitude gradually washed out, pointing towards field-induced randomization of ordered potential landscape. (**b**) Temporal fluctuations show similar suppression of fluctuations with magnetic field.

MoS$_2$/PMN-PT HC-FET with Ti/Au contact

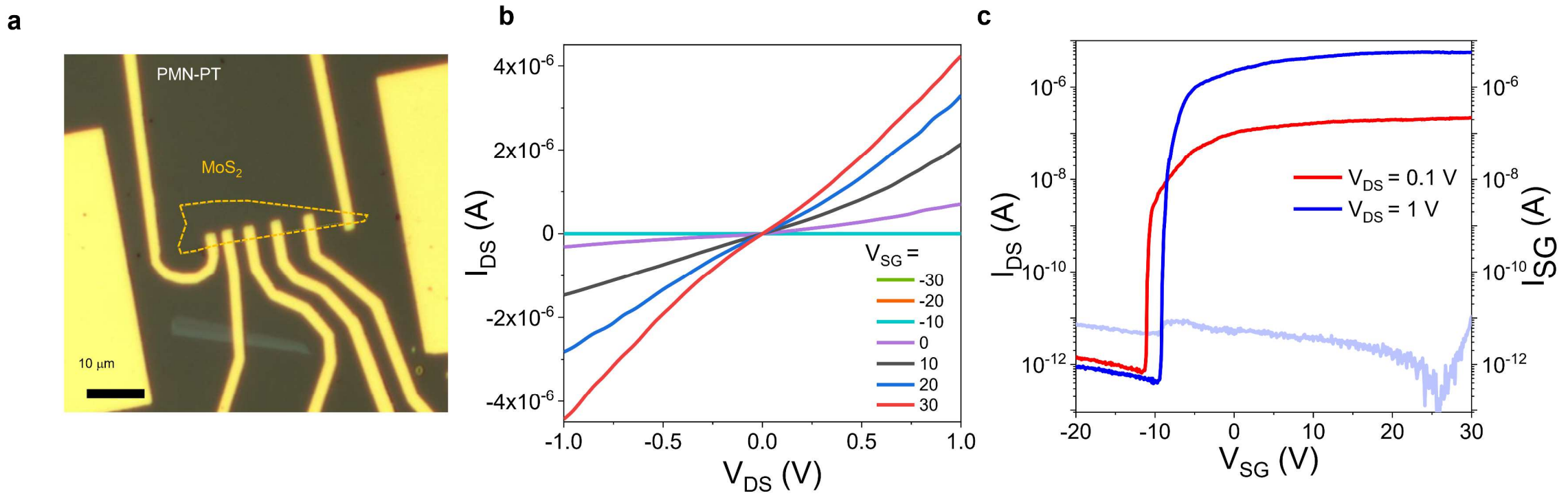

**Figure S29 | Contact-independent performance improvement in side-gate configuration**. (**a**) Optical micrograph of MoS$_2$/PMN-PT side-gate FET with Schottky type Ti/Au (10 nm/ 50nm) contacts. (**b**) Output characteristics at different gate biases. (**c**) Transfer characteristics show at least two orders of magnitude on-current density enhancement (Fig. 4D) compared to conventional FETs with Ti/Au contacts.

## Gate dependent channel E-field evolution in side-gate configuration

$SiO_2$ substrate

| $V_{DS}$ (V) | $V_{SG}$ (V) | E (MV/m) | % change in E |
|---|---|---|---|
| 1 | 0 | 0.786 | |
| 1 | 1 | 0.840 | 2.3 |
| 1 | 5 | 0.878 | 11.7 |
| 1 | 10 | 0.97 | 23.4 |

PMN-PT substrate

| $V_{DS}$ (V) | $V_{SG}$ (V) | E (MV/m) | % change in E |
|---|---|---|---|
| 1 | 0 | 0.219 | |
| 1 | 1 | 0.417 | 90.5 |
| 1 | 10 | 2.201 | 905.1 |
| 1 | 100 | 20.04 | 9051.2 |

**Table S1 |** Electric field modulation in $MoS_2$ channel in response to the applied side-gate bias. On $SiO_2$ substrate the field modulation is negligible as the side-gate bias is varied (top). In contrast, HC-FET on PMN-PT shows significant field change with the identical change in gate bias (bottom).

Simulated @ $V_{DS}$ = 1 V and $V_{gate}$ = 10 V

| Configuration | Capacitance (nF/cm$^2$) Simulated | Capacitance (nF/cm$^2$) Experimental |
|---|---|---|
| $SiO_2$ (Back gate) | 12.3 | 11.9 |
| $SiO_2$ (Side gate) | 0.48 | -- |
| PMN-PT (Side gate) | 76.3 | 40.06 |

**Table S2 |** Gate-to-channel coupling in different substrates. The simulated (FEM) capacitance values are in good agreement with the measured capacitance values for $SiO_2$ and PMN-PT indicating validity of the FEM results. Due to the low dielectric permittivity and absence of in-plane polarization, the gate-to-channel capacitive coupling is extremely poor for $SiO_2$ side-gate as evidenced from the simulated capacitance value. Meanwhile, PMN-PT shows comparable gate coupling to that of the $SiO_2$ back-gate, owing to in-plane ferroelectric polarization.

| | 2P Mobility ($cm^2 V^{-1} s^{-1}$) | SS (mV/dec) | $I_{ON}$ (μA/μm) | $I_{ON}$/ $I_{OFF}$ | Ref. |
|---|---|---|---|---|---|
| n-type HC-FET ($1L-MoS_2$) | 4636 | 68 | 153 | $10^8$ | This work |
| p-type HC-FET ($1L-WSe_2$) | 1847 | 310 | 100.84 | $10^8$ | This work |
| 1L $MoS_2$/ $SiO_2$ back-gate | 97 | 2410 | 10.22 | $>10^7$ | This work |
| $HfO_2/MoS_2/SiO_2$ | ~200 | 74 | 2.5 | $10^8$ | (*9*) |
| $MoS_2$ Bi contacts | 55 | >2000 | ~26 | $>10^7$ | (*25*) |
| 1-3L $MoS_2$/ $SiN_x$ | 236-669 | -- | -- | $10^6$ | (*26*) |
| 1-3L $MoS_2$/ $SiN_x$ | 60-820 | -- | ~16 | $10^7$ | (*27*) |
| $MoS_2$ / 1T' $MoS_2$ | 50 | 90-100 | ~70 | $>10^7$ | (*28*) |
| $MoS_2$ hBN encapsulated | 62 | -- | ~ 1 | -- | (*29*) |
| 1L $WSe_2$ strain-free transferred Pt contacts | 618 | -- | ~ 100 | $>10^8$ | (*30*) |
| 3L $MoS_2$/ $Y_2O_3$ buffer | 63.7 | -- | > 50 | $10^8$ | (*31*) |
| FL $WSe_2$ vdW Pt contacts | 190 | -- | > 20 | $10^7$ | (*32*) |
| 1L $WSe_2$ vdW Pt contacts | 10-40 | -- | ~ 10 | $<10^6$ | (*32*) |

**Table S3 |** Comparison table for key performance parameters (FET mobility, SS, maximum on-current and on-off ratio) for state-of-the-art FETs.

## References

1. Jena, D. & Konar, A. Enhancement of carrier mobility in semiconductor nanostructures by dielectric engineering. *Phys. Rev. Lett.* **98**, 136805 (2007).

2. Konar, A. & Jena, D. Tailoring the carrier mobility of semiconductor nanowires by remote dielectrics. *J. Appl. Phys.* **102**, 123705 (2007).

3. Ma, N. & Jena, D. Charge scattering and mobility in atomically thin semiconductors. *Phys. Rev. X* **4**, 011043 (2014).

4. Ong, Z. Y. & Fischetti, M. V. Charged impurity scattering in top-gated graphene nanostructures. *Phys. Rev. B* **86**, 121409 (2012).

5. Ong, Z. Y. & Fischetti, M. V. Mobility enhancement and temperature dependence in top-gated single-layer $MoS_2$. *Phys. Rev. B* **88**, 165316 (2013).

6. Utama, M. I. B. *et al.* A dielectric-defined lateral heterojunction in a monolayer semiconductor. *Nat. Electron.* **2**, 60–65 (2019).

7. Uzhansky, M. *et al.* Non-volatile reconfigurable p-n junction utilizing in-plane ferroelectricity in 2D $WSe_2$/α-$In_2Se_3$ asymmetric heterostructures. *Adv. Funct. Mater.* **34**, 2306682, (2024).

8. Lombardi, C., Manzini, S., Saporito, A. & Vanzi, M. A. Physically based mobility model for numerical simulation of nonplanar devices. *IEEE Trans. Comput. Des. Integr. Circuits Syst.* **7**, 1164–1171 (1988).

9. Radisavljevic, B., Radenovic, A., Brivio, J., Giacometti, V. & Kis, A. Single-layer $MoS_2$ transistors. *Nat. Nanotechnol.* **6**, 147–150 (2011).

10. Canali, C., Minder, R. & Ottaviani, G. Electron and hole drift velocity measurements in silicon and their empirical relation to electric field and temperature. IEEE Trans. Electron Devices 22, 1045–1047 (1975).

11. Hosseini, M., Elahi, M., Pourfath, M. & Esseni, D. strain-induced modulation of electron mobility in single-layer transition metal dichalcogenides $MX_2$ ($M$ = Mo, W; $X$ = S, Se). *IEEE Trans. Electron Devices* **62**, 3192–3198 (2015).

12. Sørensen, S. G., Füchtbauer, H. G., Tuxen, A. K., Walton, A. S. & Lauritsen, J. V. Structure and electronic properties of in situ synthesized single-layer $MoS_2$ on a gold surface. *ACS Nano* **8**, 6788–6796 (2014).

13. Velický, M. *et al.* Strain and charge doping fingerprints of the strong interaction between monolayer $MoS_2$ and gold. *J. Phys. Chem. Lett.* **11**, 6112–6118 (2020).

14. Michail, A., Delikoukos, N., Parthenios, J., Galiotis, C. & Papagelis, K. Optical detection of strain and doping inhomogeneities in single layer $MoS_2$. *Appl. Phys. Lett.* **108**, 173102 (2016).

15. Ikeda, T. *Fundamentals of Piezoelectricity*. (Oxford University Press, 1996).

16. Innovia. *Innovia Ultrasound Grade PMN-PT (001-poled) Crystal Datasheet*.

https://www.innoviamaterials.com/upLoad/down/month_2404/20240426151812120.pdf.

17. Zhao, P. *et al.* Evaluation of border traps and interface traps in $HfO_2/MoS_2$ gate stacks by capacitance-voltage analysis. *2D Mater.* **5**, 031002, (2018).

18. Zhao, P. *et al.* Probing interface defects in top-gated $MoS_2$ transistors with impedance spectroscopy. *ACS Appl. Mater. Interfaces* **9**, 24348–24356 (2017).

19. Ando, T., Fowler, A. B. & Stern, F. Electronic properties of two-dimensional systems. *Rev. Mod. Phys.* **54**, 437–672 (1982).

20. Koivisto, M. & Stott, M. J. Kinetic energy functional for a two-dimensional electron system. *Phys. Rev. B* **76**, 195103 (2007).

21. Tang, T. W., O'Regan, T. & Wu, B. Thomas-Fermi approximation for a two-dimensional electron gas at low temperatures. *J. Appl. Phys.* **95**, 7990–7997 (2004).

22. Breit, G. & Wigner, E. Capture of slow electrons. *Phys. Rev.* **49**, 519 (1936).

23. Nazarov, Y. V. & Glazman, L. I. Resonant tunneling of interacting electrons in a one-dimensional wire. *Phys. Rev. Lett.* **91**, 126804 (2003).

24. Datta, S. *Electronic Transport in Mesoscopic Systems*. (Cambridge University Press, 1995).

25. Shen, P. C. *et al.* Ultralow contact resistance between semimetal and monolayer semiconductors. *Nature* **593**, 211–217 (2021).

26. Ng, H. K. *et al.* Improving carrier mobility in two-dimensional semiconductors with rippled materials. *Nat. Electron.* **5**, 489–496 (2022).

27. Liu, T. *et al.* Crested two-dimensional transistors. *Nat. Nanotechnol.* **14**, 223–226 (2019).

28. Kappera, R. *et al.* Phase-engineered low-resistance contacts for ultrathin $MoS_2$ transistors. *Nat. Mater.* **13**, 1128–1134 (2014).

29. Liu, Y. *et al.* Toward barrier free contact to molybdenum disulfide using graphene electrodes. *Nano Lett.* **15**, 3030–3034 (2015).

30. Liu, Y. *et al.* Low-resistance metal contacts to encapsulated semiconductor monolayers with long transfer length. *Nat. Electron.* **5**, 579−585 (2022).

31. Zou, X. *et al.* Interface engineering for high-performance top-gated $MoS_2$ field-effect transistors. *Adv. Mater.* **26**, 6255–6261 (2014).

32. Wang, Y. *et al.* P-type electrical contacts for 2D transition-metal dichalcogenides. *Nature* **610**, 61–66 (2022).